\documentclass[12pt]{article}

\usepackage{pgfplots}
\usetikzlibrary{decorations.pathmorphing}

\tikzset{snake it/.style={decorate, decoration=snake}}
\pgfplotsset{compat=1.10}
\usepgfplotslibrary{fillbetween}
\usetikzlibrary{patterns}

\usepackage{draft} 
\usepackage{hyperref}
\usepackage{subfigure}
\usepackage{graphicx,color}
\usepackage{cite}
\usepackage{mciteplus}
\usepackage{skak}
\usepackage{xcolor}
\usepackage{empheq}
\usepackage{tikz}
\DeclareFontFamily{OT1}{pzc}{}
\DeclareFontShape{OT1}{pzc}{m}{it}{<-> s * [1.10] pzcmi7t}{}
\DeclareMathAlphabet{\mathpzc}{OT1}{pzc}{m}{it}

\def\be#1\ee{\begin{align}#1\end{align}}



\begin{document}

\unitlength = .8mm

\begin{titlepage}

\begin{center}

\hfill \\
\hfill \\
\vskip 1cm

\title{\Huge From Horowitz -- Polchinski  \\ to Thirring and Back}

\author{Jinwei Chu and David Kutasov}
\address{
Leinweber Institute for Theoretical Physics,\\  Enrico Fermi Institute, and Department of Physics\\ University of Chicago, Chicago IL 60637
}
\vskip 1cm

\email{jinweichu@uchicago.edu, dkutasov@uchicago.edu}

\end{center}

\abstract{We propose a new approach for studying $d+1$ dimensional Euclidean Schwarzschild black holes with Hawking temperature near the Hagedorn temperature and Horowitz-Polchinski solutions. The worldsheet theory that describes some of these backgrounds is strongly coupled. We use its underlying affine $SU(2)_L\times SU(2)_R$ symmetry to continue to weak coupling, by varying the level of the current algebra from the small value relevant for black holes and HP solutions to a large value. In this limit, one can describe the dynamics by a solvable effective field theory, and the non-geometric features of the original problem are geometrized. The resulting construction is closely related to previous work on the non-abelian Thirring model, and sheds light on both problems.
}

\vfill

\end{titlepage}

\eject

\begingroup
\hypersetup{linkcolor=black}
\tableofcontents
\endgroup

\vfill
\eject

\section{Introduction}
\label{intro}

In General Relativity (GR), a Euclidean Schwarzchild black hole in asymptotically flat $d+1$ dimensional spacetime is a solution of Einstein's equations, described by the metric 
\ie
\label{bh}
ds^2=f(r)d\tau^2+\frac{dr^2}{f(r)}+r^2d\Omega_{d-1}^2~,
\fe
where $(r, \Omega_{d-1})$ are spherical coordinates on ${\mathbb{R}}^d$,
\ie
\label{fbeta}
f(r)=1-\left(\frac{r_0}{r}\right)^{d-2}~,
\fe
and $r_0$ is the Schwarzschild radius. The Euclidean time $\tau$ lives on a circle of circumference 
\ie
\label{bbb}
\beta=\frac{4\pi r_0}{d-2}\ ,
\fe
the inverse Hawking temperature of the black hole. 

The spacetime \eqref{bh} -- \eqref{bbb} can be thought of as describing a normalizable state in a theory living in the asymptotic large $r$ geometry ${\mathbb{R}}^d\times \mathbb{S}^1$. In this state, the radial coordinate $r$ and Euclidean time $\tau$ form a semi-infinite cigar. The spacetime \eqref{bh} -- \eqref{bbb} plays an important role in black hole thermodynamics.

When we embed GR in (classical) string theory, Einstein's equations are modified by perturbative and non-perturbative $\alpha'$ corrections, and it is natural to ask what their effect is on the solution \eqref{bh} -- \eqref{bbb}. These corrections are small when $\beta$ is much larger than the string length $l_s=\sqrt{\alpha'}$, but are expected to become significant for $\beta\sim l_s$.

An important qualitative phenomenon associated with the solution \eqref{bh} -- \eqref{bbb} in string theory is the breaking of the winding symmetry around the Euclidean time circle, due to the fact that a string wound around the circle at large $r$ can move to the vicinity of the Euclidean horizon, $r=r_0$, where it becomes a standard, unwound, string propagating near the tip of the cigar. This breaking is spontaneous, since it is a feature of the particular  state \eqref{bh} -- \eqref{bbb},  and not of the asymptotic ${\mathbb{R}}^d\times \mathbb{S}^1$ background.

Another source of winding number non-conservation is an expectation value of the field that at large $r$ describes a string wound once around the Euclidean time circle. This string gives rise on ${\mathbb{R}}^d$ to a complex scalar field $\chi$. For low temperature, $\beta\gg l_s$, this field is heavy at infinity, but in the background \eqref{bh} -- \eqref{bbb} it has a non-zero expectation value, that depends on the radial coordinate, $\chi(r)$ (see e.g. \cite{Kutasov:2005rr, Chen:2021dsw} for discussions). For large $\beta$, this expectation value goes rapidly to zero as $r\to\infty$. The non-zero $\chi$ gives a non-perturbative $\alpha'$ correction to the background \eqref{bh} -- \eqref{bbb}. It can be thought of as an order parameter for the winding symmetry breaking.

As mentioned above, as $\beta$ decreases, the gravity solution \eqref{bh} -- \eqref{bbb} becomes less reliable due to $\alpha'$ corrections. In the full string theory, the GR background is replaced by a worldsheet CFT that approaches ${\mathbb{R}}^d\times \mathbb{S}^1$ at large $r$, with curvature localized in a region of size $r_0$. For $r_0\sim l_s$, this CFT is strongly coupled (on the worldsheet). Understanding it is an important open problem, for example for the string/black hole correspondence \cite{Horowitz:1996nw}, and the related problem of understanding black hole microstates. 

When the inverse Hawking temperature $\beta$ \eqref{bbb} approaches the inverse Hagedorn temperature $\beta_H=2\pi R_H$, 
\begin{equation}
	\label{tthh}
	\begin{split}
    R_H^{\rm bosonic}=2l_s\ ,\;\; R_H^{\rm type\; II}=\sqrt2l_s\ ,
	\end{split}
\end{equation}
one may try to study the EBH \eqref{bh} -- \eqref{bbb} by using an effective field theory (EFT). Since in this regime the Euclidean time circle is string size, one must reduce on this circle, and write down a $d$ dimensional effective action on ${\mathbb{R}}^d$. That action is expected to contain the radion field $\phi$, a (real) massless field that parametrizes the radius of Euclidean time, and the winding tachyon field $\chi$ mentioned above, which is light near the Hagedorn temperature. At large $r$, its mass is given by 
\ie
	\label{minfty}
    m_\infty^2=\frac{R^2-R^2_H}{\alpha'^2}~,
\fe
which goes to zero as $\beta=2\pi R\to \beta_H=2\pi R_H$. 

Horowitz and Polchinski (HP) \cite{Horowitz:1997jc} studied the effective action of $\phi$ and $\chi$ to leading non-trivial order in the fields, and looked for solutions that are normalizable and preserve the $SO(d)$ symmetry of ${\mathbb{R}}^d$. As we review in section \ref{review}, they found\footnote{See \cite{Chen:2021dsw,Balthazar:2022szl,Balthazar:2022hno,Mazel:2024alu,Bedroya:2024igb} for some recent discussions and further references.} that
for $d=3,4,5$, the HP EFT has normalizable solutions of radial size $\sim 1/m_\infty$, which is assumed to be much larger than $l_s$. These solutions only exist for non-zero $m_\infty$, i.e. slightly below the Hagedorn temperature. They go to zero for all $r$ in the limit $\beta\to\beta_H$ (or $m_\infty\to 0$). For $d=6$, the HP EFT has normalizable solutions only for $m_\infty=0$, i.e. at the Hagedorn temperature \cite{Balthazar:2022szl}. There is a continuum of such solutions, labeled by a parameter that determines their size. For $d>6$, the HP EFT has no normalizable solutions. 

The relation of the HP solutions to small EBH's (i.e. EBH's with $\beta\sim \beta_H$) is  unclear for a number of reasons. One is that if we naively continue the EBH solution \eqref{bh} -- \eqref{bbb} to the regime $\beta\sim\beta_H$, the non-trivial geometry seems to be confined to a region of radial size $\sim l_s$, whereas the HP solution is extended over a region of size $1/m_\infty\gg l_s$. Another is that the HP EFT does not have solutions for $d> 6$, whereas the EBH with $\beta\sim\beta_H$ is expected to exist for all $d$. For large $d$ this has been shown explicitly in \cite{Emparan:2013xia,Chen:2021emg}.  In that regime,  the EBH develops a throat which looks like a cut-off coset CFT $SL(2,\mathbb{R})/U(1)$, with the level of  $SL(2,\mathbb{R})$ related to the inverse temperature, $\beta$. This construction allows one to interpolate continuously between large black holes with $\beta\gg l_s$, and small black holes with $\beta\sim l_s$. 

A third qualitative difference between EBH's and HP solutions is that the former have the topology of a disk in $(\tau,r)$ space, while the latter have the topology of a cylinder. In other words, the Euclidean time circle contracts to a point at some finite $r$ for EBH's, and does not for HP solutions. Of course, for $\beta\sim l_s$ the geometry is not trustworthy but, at least in the worldsheet supersymmetric case, one may be able to distinguish different topologies by the Witten index of the corresponding worldsheet theories \cite{Chen:2021dsw}. 

In our previous work \cite{Balthazar:2022szl,Balthazar:2022hno} we took some steps towards clarifying the situation. In particular, we showed that for $d=6$, the solutions of the HP EFT, that only exist at the Hagedorn temperature, are modified qualitatively by the addition of the leading $\alpha'$ corrections to the HP Lagrangian. In the presence of these corrections, these solutions are shifted from $m_\infty=0$ to $m_\infty\not=0$. Their radial size, which in the absence of these corrections is a free parameter, becomes a function of $m_\infty$; it scales like 
$\sqrt{l_s/m_\infty}$. The resulting solutions vanish in the limit $m_\infty\to 0$, like the HP solutions for $d<6$.    

For $d>6$, the qualitative change due to adding $\alpha'$ corrections is even more pronounced. As mentioned above, without these corrections there are no normalizable solutions at all. As we review in section \ref{review}, the addition of even just the leading corrections changes that. One interesting feature of the $d>6$ solutions is that they are non-trivial {\it at} the Hagedorn temperature, i.e. at $m_\infty=0$. This is in contrast with the situation for $d\le 6$, where the solutions vanish at the Hagedorn temperature.  

For general $d>6$, the solutions of the modified HP EFT are strongly coupled, and analyzing them requires knowledge of the full effective action. However, for $d=6+\epsilon$ with $\epsilon\ll 1$, one can study them perturbatively in $\epsilon$. For example, to leading order in $\epsilon$, we found in \cite{Balthazar:2022hno} that the radial size of the solution at the Hagedorn temperature scales like $l_s/\sqrt{\epsilon}$. Thus, as $\epsilon$ increases, the size of the solution decreases. 

To study the solutions at integer values of $d$, such as $d=7, 8, \cdots$, we need to set $\epsilon$ to a positive integer value. As we will review, in the study of the HP EFT this corresponds to having to  include in the effective action terms of arbitrarily high orders in the fields $\phi$, $\chi$ (as well as the dilaton and metric) \cite{Balthazar:2022hno}. This is the sense in which this problem is strongly coupled. At the same time, a naive extrapolation of the small $\epsilon$ analysis suggests that the analogs of HP solutions for finite $\epsilon$ are localized at the string scale, like one would expect from the EBH \eqref{bh} -- \eqref{bbb} in the limit $\beta\to\beta_H$. Thus, it is natural to conjecture that the HP-type solutions for $d>6$ might be continuously connected to large EBH's, unlike their $d<6$ cousins. In this paper and the companion paper~\cite{Chu:2025kzl}, we will see that the actual situation is more interesting.

Further progress along the lines of \cite{Balthazar:2022hno} was impeded by the fact that to proceed one needs to solve a strongly coupled CFT. In the EFT approach, one needs to know the full generalized HP effective action, which is difficult to calculate. In this paper we suggest an approach that allows one to circumvent this problem. 

The starting point of our approach is the observation that at the Hagedorn temperature, the symmetry of string theory on ${\mathbb{R}}^d\times {\mathbb{S}}^1$ is enhanced from $U(1)_L\times U(1)_R$, corresponding to momentum and winding on the circle, to $SU(2)_L\times SU(2)_R$ \cite{Balthazar:2022szl}. The spacetime fields $\phi$ and $\chi$ correspond from the worldsheet point of view to couplings in a generalized non-abelian Thirring model. Thus, the problem of calculating the spacetime effective Lagrangian is equivalent to studying the worldsheet dynamics of a generalized non-abelian Thirring model for $SU(2)$, with couplings that depend on the position in ${\mathbb{R}}^d$. This is an interesting problem in its own right. It is in general difficult, but we can borrow ideas and tools developed in that context to make progress in ours. 

One idea, that will be a main theme of this paper, is the following. In the HP problem, the $SU(2)_L\times SU(2)_R$ symmetry is generated by Kac-Moody currents $J^a(z)$, $\bar J^{\bar b}(\bar z)$ at a level of order one ($k=1$ in the bosonic string, and $k=2$ in the type II theory). As we will explain, for large $k$ the problem simplifies significantly, while maintaining some essential non-trivial features. We will use this simplification to find the exact HP Lagrangian for large $k$, and try to learn from it about the features of interest for small $k$. This is in the spirit of large $N$ approximations in QFT, where it is known  that large $N$ theories often capture qualitative and sometimes even quantitative properties of theories with $N$ of order one. 

At large $k$, the $SU(2)$ WZW model describes a sigma-model on a large three-sphere, with radius $R=\sqrt{k}l_s$ \cite{Witten:1983ar}. As we will discuss later, in this limit some  non-geometric features of the original HP system become geometric. In particular, the winding tachyon, which does not have a geometric interpretation on $\mathbb{S}^1$, becomes a geometric mode on the three-sphere at large $k$. Thus, one can view the large $k$ analogs of HP solutions as ones in which the size and shape of a large three-sphere change with the radial direction on $\mathbb{R}^d$. 

In the context of non-abelian Thirring, the large $k$ approximation made an appearance already in \cite{Kutasov:1989dt}, and we will use the results of that paper in our problem. We will need a more general version of the calculations in \cite{Kutasov:1989dt}, which are useful in that context as well. In particular, \cite{Kutasov:1989dt} considered the case where the Thirring interaction preserves an  $SU(2)_{\rm diag}\subset SU(2)_L\times SU(2)_R$, while for our purposes this restriction needs (in general) to be relaxed.   

An important point is that while in the HP context, $SU(2)_L\times SU(2)_R$ is only a symmetry {\it at} $\beta=\beta_H$, it is useful away from the Hagedorn temperature as well. The reason is that from the spacetime point of view, the breaking of the symmetry $SU(2)_L\times SU(2)_R\to U(1)_L\times U(1)_R\to U(1)_{\rm diag}$ for $\beta\not=\beta_H$ is spontaneous. Thus, the spacetime Lagrangian, which we will compute at large $k$, must have the larger symmetry, and it is only the expectation values of the fields that break it to the lower one. This is of course a standard situation in QFT, which helps analyze theories with spontaneous symmetry breaking.

The plan of this paper is the following. In section \ref{review} we review some prior results that will play a role in our discussion. We start with a discussion  of the HP EFT and its solutions for $d< 6$. As mentioned above, these solutions only exist for temperatures below the Hagedorn temperature. Their radial size is proportional to the Compton wavelength of the winding tachyon $\chi$, $1/m_\infty$ \eqref{minfty}. In particular, it diverges in the limit $\beta\to\beta_H$. For $d=6$, solutions only exist at the Hagedorn temperature. Their size is a free parameter; different solutions are related by a scaling symmetry of the equations. For $d>6$, the HP EFT does not have normalizable solutions. 

We also review the results of \cite{Balthazar:2022hno} on the role of higher order corrections to the HP effective Lagrangian. For $d<6$, these corrections have only minor effects (near the Hagedorn temperature), but for $d\ge 6$ they qualitatively change the picture. In particular, for $d=6$ the solution goes from existing only at the Hagedorn temperature, to only existing below it. For $d>6$, the modified HP EFT has normalizable solutions for all $\beta\ge \beta_H$, in contrast to the original one, that does not have such solutions. An important role in our discussion is played by the $SU(2)_L\times SU(2)_R$ symmetry of the small EBH system \cite{Balthazar:2022szl}. We review the construction of this symmetry, and its implications  for the effective action. 

In section \ref{seclargek} we present the problem we are interested in -- the extension of the HP solutions to $d> 6$. The strongly coupled nature of this problem leads us to seek solvable approximations. We describe such an approximation -- the extension of the $SU(2)_L\times SU(2)_R$ affine Lie algebra of the theory from level $k$ of order one in the original HP system, to large $k$. We show that in this limit, the effective Lagrangian simplifies significantly, while retaining the kind of non-trivial structure that we expect to find at small $k$ as well. 

The resulting Lagrangian takes the form \eqref{efflagdil}. It consists of dilaton gravity coupled to a two derivative Lagrangian. In sections \ref{metric} and \ref{potential} we calculate the kinetic and potential terms in this Lagrangian, respectively. We check that our result agrees with the gravity analysis~\cite{Cvetic:2000dm}. We describe the normalizable solutions of the equations of motion of this Lagrangian in a companion paper \cite{Chu:2025kzl}. 

In section \ref{beyond} we go back to the original HP system (i.e. to $k$ of order one), and consider the question what we can say about the structure of the generalized HP effective action using the $SU(2)_L\times SU(2)_R$ symmetry and other considerations. In section \ref{discuss} we briefly discuss our results and possible extensions. In particular, we argue that in addition to the backgrounds studied in this paper, that are large $k$ analogs of HP solutions, there may be a second class of backgrounds, that are the analogs of small black holes. These other backgrounds involve non-trivial condensates of spherical harmonics on the three-sphere in \eqref{wzw}, which are set to zero in the analysis of this paper. The existence of these backgrounds is motivated by analogy to the study of systems of NS5-branes. We leave their construction to future work.

\section{Review}
\label{review}

In this section we will review some prior work that will play a role in our discussion below. 

\subsection{HP EFT}

We start with the Euclidean spacetime ${\mathbb{R}}^d\times {\mathbb{S}}^1$, and take the radius of the ${\mathbb{S}}^1$ to be the Hagedorn radius $R_H$ \eqref{tthh}. As mentioned in section \ref{intro}, this choice will allow us to use the enhanced $SU(2)_L\times SU(2)_R$ symmetry that appears at this value of the radius, but the analysis will be applicable to general $R$ as well.

We focus on the dynamics of two fields,\footnote{As explained in \cite{Chen:2021dsw,Balthazar:2022hno}, the dilaton and metric can be neglected to leading order in $\phi$, $\chi$.} the radion $\phi(x)$ and winding tachyon $\chi(x)$, where $x$ labels position in ${\mathbb{R}}^d$. $\phi(x)$ parametrizes the local radius of the ${\mathbb{S}}^1$,
\ie
\label{rx}
R(x)=R_H \left[1+\phi(x)+O(\phi^2)\right]~.
\fe
The higher order terms in $\phi$ in \eqref{rx} depend on the choice of coordinates on $\phi$ space, or equivalently on the choice of contact terms in the worldsheet CFT  \cite{Kutasov:1988xb}. We will return to this issue below. The field $\phi$ is massless and has a flat potential. Indeed, any constant $\phi$ leads to a CFT -- varying $\phi$ corresponds to varying the radius of the ${\mathbb{S}}^1$, \eqref{rx}. 

Turning to the winding tachyon $\chi$, for general $R$ this field has mass $m_\chi$. Locally in ${\mathbb{R}}^d$ it is given by a formula similar to \eqref{minfty}, namely
\begin{equation}
\label{localmass}
    m_\chi^2=\frac{R^2(x)-R_H^2}{\alpha'^2}\ .
\end{equation}
$m_\infty$ in equation \eqref{minfty} is the asymptotic mass of $\chi$, $m_\infty=\lim_{x\to\infty} m_\chi(x)$. It is determined by the asymptotic value of $R$ or, equivalently \eqref{rx}, of $\phi$. As mentioned above, it vanishes at $R=R_H$, or $\phi=0$. 

To leading order in the fields, the Lagrangian for $\phi$ and $\chi$ takes the form 
\begin{equation}
	\label{lsix}
	\begin{split}
L_6=(\nabla\phi)^2+|\nabla \chi|^2+\frac{2R^2_H}{\alpha'^2}\phi|\chi|^2\ .
	\end{split}
	\end{equation}
A few comments about this Lagrangian:
\begin{itemize}
\item We omitted an overall multiplicative factor in \eqref{lsix}. It  will not play an important role in our discussion, but the value in \eqref{lsix} will be useful below.
\item We normalized $\chi$ such that its kinetic term in \eqref{lsix} has the same coefficient as that for $\phi$. This turns out to be convenient later, when we talk about the role of the $SU(2)_L\times SU(2)_R$ symmetry.
\item The coefficient of the cubic term in the Lagrangian is fixed by the requirement that the mass of $\chi$ comes out correctly when substituting  \eqref{rx} in \eqref{localmass}. 
\item The Lagrangian \eqref{lsix} has a scaling symmetry under which $x$ has dimension $-1$, and $\phi$, $\chi$ have dimension two. Under this symmetry the Lagrangian has dimension six, which is the reason for the subscript in \eqref{lsix}. Corrections to \eqref{lsix} involve operators of higher dimension w.r.t. this scaling symmetry.
\item In this paper we will mainly discuss  the bosonic string, in which $R_H=2l_s$ \eqref{tthh}, but in \eqref{lsix} we left it free, since the discussion can be generalized to the superstring. Some aspects of this generalization are discussed in \cite{Balthazar:2022szl}. 
\end{itemize}

The equations of motion of the Lagrangian \eqref{lsix} are
\begin{equation}
	\label{eomsix}
	\begin{split}
 	\nabla^2\chi &=\frac{2R^2_H}{\alpha'^2}\phi\chi \ ,\\
 	\nabla^2\phi &=\frac{R_H^2}{\alpha'^2}|\chi|^2\ .
	\end{split}
	\end{equation}
These equations were studied in the past in a different context, e.g. in \cite{Ruffini:1969qy,Friedberg:1986tp,Moroz:1998dh}, and in the HP context in \cite{Horowitz:1997jc,Chen:2021dsw,Balthazar:2022szl,Balthazar:2022hno,Bedroya:2024igb}. We are looking for normalizable solutions that preserve the $SO(d)$ symmetry, i.e. $\chi=\chi(r)$, $\phi=\phi(r)$. We  also take $\chi(r)$ to be real and positive (w.l.g.). Below the Hagedorn temperature, the solutions behave at large $r$ like
\begin{equation}
    \begin{split}
        &\chi(r)\sim A_\chi r^{-\frac{d-1}{2}}e^{-\frac{\sqrt{2\phi_{\infty}}R_H}{\alpha'} r}\ ,
        \\
        &\phi(r)\sim \phi_\infty+C_\phi r^{-(d-2)}\ ,
    \end{split}
    \label{chiphilarger}
\end{equation}
with some constants $A_\chi, C_\phi$. 

The constant $\phi_\infty=\lim_{r\to\infty} \phi(r)$  in \eqref{chiphilarger} determines the asymptotic radius of $\mathbb{S}^1$ via (\ref{rx}). It is positive for temperatures below the Hagedorn temperature, and is related to $m_\infty$ \eqref{minfty} via
\begin{equation}
\label{phim}
m_\infty=\frac{\sqrt{2\phi_{\infty}}R_H}{\alpha'}\ .
\end{equation}
For $\phi_\infty>0$ (i.e. below the Hagedorn temperature), $\chi(r)$ \eqref{chiphilarger} decays exponentially, like $\exp(-rm_\infty)$, and thus is normalizable. Regularity at $r=0$  leads to the conditions $\chi'(0)=\phi'(0)=0$. Then, integrating both sides of the second equation in (\ref{eomsix}) from 0 to any finite $r$, we find that $\phi'(r)$ is non-negative for all $r$. This implies that $C_\phi<0$.

\begin{figure}
	\centering
	\subfigure[]{
	\begin{minipage}[t]{0.45\linewidth}
	\centering
	\includegraphics[width=2.5in]{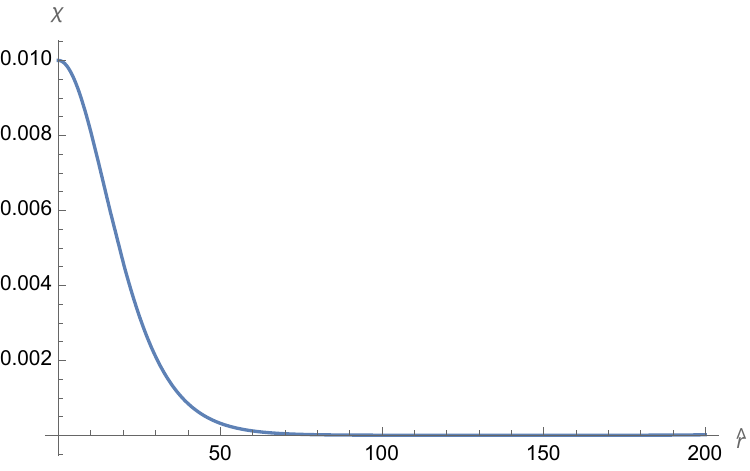}\label{chid3chi001}
	\end{minipage}}
	\subfigure[]{
	\begin{minipage}[t]{0.45\linewidth}
	\centering
	\includegraphics[width=2.5in]{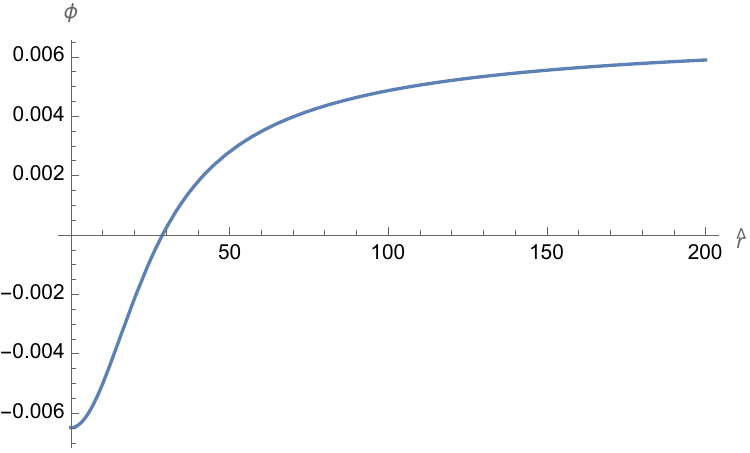}\label{phid3chi001}
	\end{minipage}}
	\centering
\caption{\label{d3chi001}The profiles of $\chi$ and $\phi$ for $d=3$ and $\chi(0)=0.01$.}
\end{figure}

In terms of the dimensionless radial coordinate $\hat r=\frac{R_H}{\alpha'} r$, (\ref{eomsix}) takes the form
\begin{equation}
\label{eomsixres}
	\begin{split}
 	\partial^2_{\hat{r}}\chi+\frac{d-1}{\hat{r}}\partial_{\hat{r}}\chi &=2\phi\chi \ ,\\
 	\partial^2_{\hat{r}}\phi+\frac{d-1}{\hat{r}}\partial_{\hat{r}}\phi&=\chi^2\ .
	\end{split}
	\end{equation}
Equations \eqref{eomsixres} do not seem to be analytically solvable, but one can solve them numerically. For example, figure \ref{d3chi001} shows a plot of $\chi(\hat r)$ and $\phi(\hat r)$ for $d=3$. To obtain this numerical solution, we chose the initial conditions $\chi(0)=0.01$, $\chi'(0)=\phi'(0)=0$, and tuned $\phi(0)$ such that $\chi(\hat r)$ goes to zero at large $\hat r$. Figure \ref{phid3chi001} shows that $\phi_\infty$ \eqref{chiphilarger} is positive in this case. This implies, via \eqref{rx}, that the solution of figure \ref{d3chi001} corresponds to a particular $R>R_H$, i.e. to a temperature slightly below the Hagedorn temperature. 

While figure \ref{d3chi001} corresponds to a particular choice of $R$, or $\beta$, one can use it to generate solutions for arbitrary $\beta$ in the regime of validity of the effective Lagrangian \eqref{lsix}, $m_\infty\ll m_s$, by utilizing the scaling symmetry of the Lagrangian \eqref{lsix} mentioned above. That symmetry implies that if $(\phi(\hat r),\chi(\hat r))$ is a solution of \eqref{eomsixres}, so is  
\begin{equation}
\label{res}
  \left(\chi_q(\hat r),\phi_q(\hat r)\right)= \left(q^2\chi(q\hat r)\ ,\ q^2\phi(q\hat r)\right)
\end{equation}
for any positive constant $q$. This rescaling takes $\chi(0)\to q^2\chi(0)$, $\phi_\infty\to q^2\phi_\infty$, and via (\ref{phim}), $m_\infty\to qm_\infty$.  

\noindent
Two immediate consequences of the symmetry \eqref{res} are:
\begin{itemize}
\item For a given value of $m_\infty$ \eqref{minfty}, \eqref{phim}, the radial size of the solution is of order $1/m_\infty$, and its height at the maximum is $\sim m_\infty^2$. 
\item In the limit $m_\infty\to 0$, which is equivalent to $q\to 0$ in \eqref{res}, the solution goes to zero for all $r$. Thus, HP solutions only exist below the Hagedorn temperature. 
\end{itemize}

\noindent
For $d=4,5$, the HP solutions behave in a qualitatively similar way to those for $d=3$. For $d=6$, one finds a different behavior. There are no solutions for $m_\infty>0$ (below the Hagedorn temperature), while for $m_\infty=0$ (at the Hagedorn temperature) there is a family of analytic solutions labeled by $\chi(0)$. These solutions take the form~\cite{Balthazar:2022szl}
\begin{equation}
\label{d6HP}
    \chi(\hat r)=-\sqrt{2}\phi(\hat r)=\frac{\chi(0)}{\left(1+\frac{\chi(0)\hat r^2}{12\sqrt{2}}\right)^2}\ ,
\end{equation}
where $\chi(0)$ is any positive number. The different solutions are related by the symmetry \eqref{res}. 

For $d>6$, there are no normalizable solutions to (\ref{eomsixres}). In particular, one can show that any bounded solution has the property that $\chi(\hat r)=-\sqrt{2}\phi(\hat r)$ and behaves like $1/\hat r^2$ at large $\hat r$, which makes the solution non-normalizable.\footnote{Solutions with these properties were constructed and studied in \cite{Bedroya:2024igb}.} The situation for $d\ge 6$ changes when higher order corrections to \eqref{lsix} are taken into account, as we describe next.

\subsection{Higher order contributions to the effective action}

As mentioned in the previous subsection, one can organize the corrections to the Lagrangian \eqref{lsix} by their dimension under the scaling symmetry described after that equation. The leading corrections have dimension eight. At that value of the dimension we can write operators with four derivatives, such as $(\nabla^2\phi)^2$, two derivatives, e.g.  $\phi(\nabla\phi)^2$, and potential terms, of which there are two,\footnote{There is no $\phi^4$ term, since for $\chi=0$, $\phi$ has a flat potential.} $\phi^2|\chi|^2$ and $|\chi|^4$. 

The four derivative terms are known to be absent in string theory. The two derivative term above can be thought of as a linear (in $\phi$) contribution to the metric on $\phi$ space. One can choose a parametrization of that space for which it vanishes. As we will see later, the $SU(2)_L\times SU(2)_R$ symmetry leads naturally to such a parametrization. We will also see that this symmetry sets to zero the coefficient of $\phi|\nabla\chi|^2$. 

This leaves the two potential terms. In \cite{Balthazar:2022hno} we wrote the combination 
\begin{equation}
\label{Leight}
    L_8=\frac{2R_H^2}{\alpha'^2}\phi^2|\chi|^2+\frac{R_H^2}{2\alpha'^2}|\chi|^4\ ,
\end{equation}
based on string amplitude calculations~\cite{Brustein:2021ifl}. As we will see later, the relative coefficient between the two terms is determined by the $SU(2)_L\times SU(2)_R$ symmetry. The overall coefficient is not qualitatively important for our discussion, though we will comment on it as well. For now, we note that looking back at \eqref{minfty}, \eqref{rx}, this coefficient implies a parametrization of the radius $R$ in terms of $\phi$,   
\begin{equation}
   R=R_H \left[1+\phi+\frac{\phi^2}{2}+O(\phi^3)\right]\ .
\end{equation}

The equations of motion of the Lagrangian $L_6+L_8$ \eqref{lsix}, \eqref{Leight} are
\begin{equation}
\label{eom68}
	\begin{split}
   	\nabla^2\chi &=\frac{2R_H^2}{\alpha'^2}\phi \chi+\frac{R_H^2}{\alpha'^2}\chi^2\chi^*+\frac{2R_H^2}{\alpha'^2} \phi^2\chi \ ,\\
 	\nabla^2\phi &=\frac{R_H^2}{\alpha'^2}|\chi|^2+\frac{2R_H^2}{\alpha'^2}|\chi|^2\phi\ .
	\end{split}
	\end{equation}
To understand the effect of the terms cubic in the fields on the r.h.s. of \eqref{eom68}, which come from (\ref{Leight}), one can proceed as follows. As we saw in the previous subsection, one consequence of the scaling symmetry \eqref{res} of the Lagrangian \eqref{lsix} is the linear relation $\phi_\infty\propto\chi(0)$. For $d=3,4,5$, the numerical results exhibited in the previous subsection imply that the proportionality coefficient in this relation is non-zero. On the other hand, for $d=6$ the analytic solution (\ref{d6HP}) has the property that $\phi_\infty=0$ for all $\chi(0)$. Therefore, in this case the  proportionality coefficient vanishes. 

The inclusion of (\ref{Leight}) in the effective Lagrangian breaks the scaling symmetry. Consequently, the above relation receives corrections, i.e. it takes the form 
\begin{equation}
\label{phiinfchi0}
    \phi_\infty=a_d\chi(0)+b_d(\chi(0))^2+\cdots\ ,
\end{equation}
where $a_d$, $b_d$ are dimension dependent constants. For $d=3,4,5$, $a_d\not=0$, so the quadratic correction in \eqref{phiinfchi0} is negligible for small enough $\chi(0)$, and the solutions of \eqref{eomsixres} are modified by a small amount due to the inclusion of the dimension eight contribution \eqref{Leight}. On the other hand, for  $d=6$, $a_6=0$, as mentioned above, and the qualitative structure of this relation changes, to $\phi_\infty= b_6(\chi(0))^2+O\left((\chi(0))^3\right)$. In other words, the corrections from (\ref{Leight}) become significant. 

The precise form of the relation \eqref{phiinfchi0} was derived in~\cite{Balthazar:2022hno}, to leading order in $\phi_\infty$ and $d-6$, namely
\ie
\label{minfchi0}
\phi_\infty=\frac{\sqrt{2}}{80}(6-d)\chi(0)+\frac{3}{140}(\chi(0))^2\ .
\fe 
This relation has the following consequences for $\epsilon=|d-6|\ll1$:
\begin{itemize}
\item For $d=6-\epsilon$, \eqref{minfchi0} implies that there are two different regimes. For $\chi(0)\ll \epsilon$, the situation is similar to that in the previous subsection, i.e. $\chi(0)\sim \phi_\infty/\epsilon $ (or $m_\infty^2/\epsilon$ in string units), and the contribution of the second term on the r.h.s. of \eqref{minfchi0} (which is due to the addition of $L_8$ \eqref{Leight} to the Lagrangian) is negligible. On the other hand, for $\epsilon\ll\chi(0)\ll 1$, the second term in \eqref{minfchi0} dominates, and one finds a linear relation $\chi(0)\sim m_\infty$. The transition between the two regimes happens at $\chi(0)\sim\epsilon$, or equivalently at $m_\infty\simeq\epsilon$.
\item For $d=6$, the qualitative nature of the solutions changes significantly when we add to the Lagrangian the dimension eight terms \eqref{Leight}. In the absence of the term quadratic in $\chi(0)$ on the r.h.s. of \eqref{minfchi0}, this equation gives $\phi_\infty=0$ for any $\chi(0)$, i.e. solutions only exist at the Hagedorn temperature. These are the solutions of \eqref{eomsixres} given by \eqref{d6HP}. However, with the quadratic term present one finds a linear relation $\chi(0)\sim m_\infty$. Thus, instead of solutions only existing {\it at} the Hagedorn temperature, we find that they only exist {\it below} it, and instead of their size being a free parameter, it is determined by the temperature (through $m_\infty$). 
\item For $d=6+\epsilon$, the quadratic term on the r.h.s. of \eqref{minfchi0} again has a dramatic effect. Without it there are no solutions for $\beta\ge \beta_H$, while with it such solutions exist. A new feature in this case compared to $d\le 6$ is the existence of a non-trivial solution {\it at} the Hagedorn temperature. Indeed, setting $\phi_\infty=0$, eq. \eqref{minfchi0} has a solution $\chi(0)\sim\epsilon$. It was shown in~\cite{Balthazar:2022hno} that this solution satisfies $\chi(r)=-\sqrt2\phi(r)$ for all $r$, a fact that will play a role in our discussion below.  
\end{itemize}

\noindent
The solutions that satisfy \eqref{minfchi0} are well approximated by the shape \eqref{d6HP} in a wide range of $r$. Thus, many of their features can be read off from this shape. In particular, the radial size of the solutions, $\Delta$, is given by\footnote{One might think that the width is still given by $ l_s/\sqrt{\phi_\infty}$, (\ref{chiphilarger}). However, (\ref{minfchi0}) implies that for $|d-6|,\phi_\infty\ll 1$, $\chi(0)\gg \phi_\infty$. This means that the region where the solution behaves as (\ref{chiphilarger}), namely $r\gg l_s /\sqrt{\phi_\infty}$, is far outside $\Delta$ (\ref{sizesol}).} 
\begin{equation}
\label{sizesol}
\Delta\sim \frac{l_s}{\sqrt{\chi(0)}}~.
\end{equation}
For example, the solutions at the Hagedorn temperature for $d=6+\epsilon$ have radial size $l_s/\sqrt\epsilon$. For small $\epsilon$ they are wide (i.e. slowly varying in $r$), which is the reason for our ability to ignore the higher order corrections to the Lagrangian $L_6+L_8$, \eqref{lsix}, \eqref{Leight}, in studying them. 

As $\epsilon$ increases, the size \eqref{sizesol} decreases, and formally substituting $\epsilon=1$ (the smallest value that gives an integer dimension) we find $\Delta\sim l_s$. In this case, the spatial extent of the solution is not large, and one expects to have to include contributions to the effective Lagrangian of terms of all dimensions  to study it. Calculating these terms is a hard problem, but in the next sections we will propose an approach to circumventing it. This approach utilizes the $SU(2)_L\times SU(2)_R$ symmetry that underlies this problem. Therefore, we next review the role of this symmetry in the HP system. As mentioned in section \ref{intro}, we will restrict our discussion to the bosonic string. The generalization to the superstring is straightforward (see \cite{Balthazar:2022szl} for a discussion).

\subsection{$SU(2)_L\times SU(2)_R$}\label{secSU2}

Renaming the Euclidean time coordinate $\tau$ in \eqref{bh} as $X$, at a general inverse temperature $\beta=2\pi R$, the asymptotic, large $r$, geometry is ${\mathbb{R}}^d\times \mathbb{S}^1$, where the $\mathbb{S}^1$ labeled by $X$ lives on a circle of radius $R$. For general $R$, the $X$ CFT has symmetry $U(1)_L\times U(1)_R$, corresponding to momentum and winding on the $X$ circle. Temperatures below the Hagedorn temperature correspond to $R>R_H$ \eqref{tthh}. 

As is well known (see e.g. \cite{Polchinski:1998rq}), at the self-dual radius, $R=l_s$, the $U(1)_L\times U(1)_R$ symmetry is enhanced to $SU(2)_L\times SU(2)_R$. The generators of $SU(2)_L$ are given by\footnote{Here $X_L$ is normalized as in \cite{Polchinski:1998rq}, $X_L(z)X_L(w)=-\frac{\alpha'}{2}\ln(z-w)$.} 
\begin{equation}
	\label{su2l}
	\begin{split}
 	J^3 &=\frac{i}{l_s}\partial X_L \ ,\\
 	J^\pm &=e^{\pm2iX_L/l_s}~ .
	\end{split}
	\end{equation}
The currents \eqref{su2l} satisfy the $SU(2)$ current algebra,
\begin{equation}
\label{su2k}
    J^a(z)J^b(0)\sim \frac{\frac{k}{2}\delta^{ab}}{z^2}+\frac{i{\epsilon^{ab}}_c}{z}J^c(0)\ ,
\end{equation}
or in components
\begin{equation}
	\label{compsu2l}
	\begin{split}
 	J^3(z)J^3(0) &\sim\frac{k/2}{z^2} \ ,\\
 	J^3(z)J^\pm(0) &\sim\frac{\pm J^\pm(0)}{z}\ ,\\
    J^+(z)J^-(0) &\sim\frac{k}{z^2}+\frac{2J^3(0)}{z}\ .
	\end{split}
	\end{equation}
\noindent
Here $k$ is the level of the affine Lie algebra \eqref{su2k}. It is equal to one for \eqref{su2l}, but in \eqref{compsu2l} we keep it general for future use. ${\epsilon^{ab}}_c$ are the structure constants of $SU(2)$. There is a similar right-moving $SU(2)$ current algebra constructed out of the right-moving components of $X$, $X_R$. 

The theory at the Hagedorn radius $R=R_H=2l_s$ can be thought of as a $\mathbb{Z}_2$ orbifold of the one at the self-dual radius. Under this $\mathbb{Z}_2$, the charged $SU(2)$ generators $J^\pm$, $\bar J^\pm$ are odd, and are therefore projected out of the theory. However, we are interested in bilinears in these operators that are invariant under the $\mathbb{Z}_2$. In particular, as is standard in string theory, from the worldsheet point of view, the spacetime fields $\phi(x)$ and $\chi(x)$ correspond to deformations of the worldsheet Lagrangian of the form \cite{Balthazar:2022szl}
\begin{equation}
\label{lws}
   {\cal L}_{\rm int}=-2\phi(x)J^3\bar J^3+\frac{1}{\sqrt2}\chi(x)J^+\bar J^-+\frac{1}{\sqrt2}\chi^*(x)J^-\bar J^+\ .
\end{equation}
Indeed, plugging (\ref{su2l}) and its right-moving analog into (\ref{lws}), we see that $\phi(x)$ multiplies $\partial X\bar\partial X$ and therefore can be thought of as a local (in $\mathbb{R}^d$) deformation of the radius of the $X$ circle, while $\chi(x)$ multiplies the vertex operator of the winding one tachyon. The normalizations in \eqref{lws} were chosen such that the worldsheet operators multiplying the real scalar fields $\phi(x)$, ${\rm Re}\;\chi(x)$, and ${\rm Im}\;\chi(x)$ are normalized to one. 

Thus, the problem of constructing HP solutions is equivalent to the worldsheet problem of finding zeroes of the $\beta$-function of the sigma model \eqref{lws}, in which the asymptotic value of $\phi$, $\phi_\infty$ \eqref{chiphilarger}, is small. Of course, \eqref{lws} is in general not the full Lagrangian, since one has to include the back-reaction of the dilaton and metric on ${\mathbb{R}}^d$ to turning on non-zero $\phi(x)$, $\chi(x)$. We will discuss this back-reaction later.

One of the interesting consequences of rephrasing the HP problem in this way is that it sheds some light on features of the spacetime effective action that are otherwise mysterious. An example is the solutions at $\beta=\beta_H$, $d>6$, which we discussed in the previous subsection. We mentioned there that these solutions satisfy the constraint $\chi(r)=-\sqrt2\phi(r)$ for all $r$. This is related to the fact that the two e.o.m. \eqref{eom68} become identical when this constraint is imposed. 

From the EFT perspective this seems mysterious, and in particular one may wonder whether this phenomenon persists in the presence of higher order contributions to the effective Lagrangian. The worldsheet description \eqref{lws} makes it clear what is going on, and suggests that the equality $\chi=-\sqrt2\phi$ for $\beta=\beta_H$ is exact. Indeed, substituting this constraint into \eqref{lws} gives 
\begin{equation}
\label{lwssym}
    {\cal L}_{\rm int}=-\phi(x)k_{a\bar b}J^a\bar J^{\bar b}~,
\end{equation}
where $k_{a\bar b}$ is a diagonal matrix, with $k_{33}=2$, $k_{+-}=k_{-+}=1$. This matrix is the inverse of the matrix of two point functions $k^{ab}=\langle J^aJ^b\rangle$. Thus, the perturbation \eqref{lwssym} preserves a diagonal $SU(2)\subset SU(2)_L\times SU(2)_R$. One of the results of \cite{Balthazar:2022hno} was that for $d>6$ the HP-type solutions discussed in the previous subsection have the property that at the Hagedorn temperature they preserve an $SU(2)_{\rm diag}$ subgroup of $SU(2)_L\times SU(2)_R$, in contrast to the solutions below the Hagedorn temperature, that only preserve a $U(1)_{\rm diag}$. This is expected to be an exact feature of the theory, and thus should persist to all orders in the perturbative expansion described in the previous subsections. 

A related feature of the worldsheet presentation \eqref{lws} is that the effective Lagrangian for $\phi$, $\chi$ (as well as the dilaton and metric) should be invariant under $SU(2)_L\times SU(2)_R$. In general this symmetry is broken by the solution $(\phi(r),\chi(r))$, but this breaking is spontaneous. 

Thus, one can phrase the generalized HP problem as follows. We add to the worldsheet Lagrangian of the free theory on ${\mathbb{R}}^d\times \mathbb{S}^1$ the perturbation 
\begin{equation}
\label{lwsgen}
    {\cal L}_{\rm int}=-\frac{2}{k}\phi_{a\bar b}(x)J^a\bar J^{\bar b}~,
\end{equation}
and write down the corresponding spacetime effective Lagrangian on ${\mathbb{R}}^d$. This Lagrangian is $SU(2)_L\times SU(2)_R$ invariant, by construction, with the fields $\phi_{a\bar 
b}$ transforming as $(3,3)$ under $SU(2)_L\times SU(2)_R$. For example, 
the kinetic term in \eqref{lsix} can be written in terms of $\phi_{a\bar b}$ (up to an overall constant) as
\begin{equation}
\label{kinterm}
\nabla\phi_{a\bar b}\nabla\phi^{a\bar b}
\end{equation}
where the $SU(2)_L$ index $a$ is raised with the $SU(2)_L$ invariant metric $\delta^{ab}$, and similarly for $SU(2)_R$. The cubic interaction in \eqref{lsix} is proportional to 
\begin{equation}
\label{cubicL}
\det\phi_{a\bar b}=\frac16\epsilon^{a_1a_2a_3}\epsilon^{\bar b_1\bar b_2\bar b_3}\phi_{a_1\bar b_1}\phi_{a_2\bar b_2}\phi_{a_3\bar b_3}
\end{equation}
which is manifestly $SU(2)_L\times SU(2)_R$ invariant. We will discuss the analogous structure of the higher order terms later in the paper. 

Since the spacetime Lagrangian is invariant under $SU(2)_L\times SU(2)_R$, the e.o.m. for the fields $\phi_{a\bar b}$ are covariant. The generalized HP solutions correspond to solutions of these equations for which we take the matrix $\phi_{a\bar b}$ to be the form (\ref{lws}). Since we are looking for $SO(d)$ invariant solutions, the functions $\phi_{a\bar b}$ can only depend on the radial coordinate in ${\mathbb{R}}^d$, $\phi_{a\bar b}=\phi_{a\bar b}(r)$. 

The above presentation of the HP problem makes it clear that it is closely related to the problem of calculating the $\beta$-function of a generalized non-abelian Thirring model for $SU(2)$. This is a difficult problem, which in our case is further complicated by the fact that the Thirring couplings $\phi_{a\bar b}$ depend on the radial coordinate on ${\mathbb{R}}^d$, $r$. 
However, in the next sections we will propose an approach to this problem which is inspired by what was done in the Thirring context in \cite{Kutasov:1989dt}. In the process, we will also shed some light on that problem.  

\section{The large $k$ limit}\label{seclargek}

As reviewed in section \ref{review}, to understand the analogs of the HP solutions for $d>6$, we need to solve a problem that can be formulated as follows. We start with the worldsheet CFT 
\begin{equation}
\label{wzw}
{\mathbb{R}}^d\times SU(2)_k/\mathbb{Z}_2~,
\end{equation}
where the second factor stands for $SU(2)$ WZW CFT with level $k$, modded by the $\mathbb{Z}_2$ symmetry described in section \ref{review}. The level $k$ is equal to one for the bosonic string. 

We introduce a perturbation of the CFT \eqref{wzw}, that takes the form \eqref{lwsgen}, and look for fixed points of the resulting worldsheet theory that preserve the $SO(d)$ rotation symmetry of \eqref{wzw}, and the $U(1)$ generated by $J_3+\bar J_3$, which corresponds to the conserved momentum around the Euclidean time circle. In the language of \eqref{lwsgen} this means that we are looking for $\phi_{a\bar b}$ of the form \eqref{lws}. 

From the spacetime point of view, this problem can be described as follows. The worldsheet couplings $\phi_{a\bar b}$ in \eqref{lws}, \eqref{lwsgen} correspond to massless fields in spacetime. One needs to write the effective Lagrangian for these fields, and look for solutions of their equations of motion.
What makes this problem difficult is that, as we saw in section \ref{review}, for general $d$ the radial size of potential solutions in ${\mathbb{R}}^d$ is expected to be $\sim l_s$. This means that in the effective action for $\phi_{a\bar b}$ we need to keep terms of all orders in fields and derivatives. In other words, there is no small parameter we can rely on. This is in contrast to the HP solutions in $d\le 6$, which have fields that scale like $m_\infty^2$, and radial size that scales like $1/m_\infty$. This allows us to neglect terms of high orders in the fields and number of derivatives. 

Calculating terms of high order in the fields in the spacetime effective Lagrangian is difficult. To see why, consider e.g. the $|\chi|^4$ term in \eqref{Leight}. To calculate this term we need to start with the scattering amplitude of four winding tachyons, and send their momenta to zero. This limit is singular, since it includes contributions from exchange of radions due to the cubic interaction in \eqref{lsix}. These contributions (as well as those of intermediate gravitons/dilatons) need to be subtracted before taking the zero momentum limit. For the $|\chi|^4$ case this calculation is tractable (see e.g. \cite{Brustein:2021ifl}), but its complexity rapidly increases with the number of fields.  

From the worldsheet point of view, this complexity is reflected in the calculation of the $\beta$-function of the non-abelian Thirring model. The way that calculation normally proceeds is by calculating correlation functions in the presence of the interaction \eqref{lwsgen}, looking for logarithmically divergent terms, and using the Callan-Symanzik equation. At high orders in the couplings, there are multiple divergences that are due to lower order terms in the $\beta$-function, that need to be subtracted to calculate the ``new" divergences that are due to higher order terms in the couplings.

As mentioned above, the basic difficulty is that the problem we are interested in does not have a small parameter that can simplify the calculation -- it is strongly coupled on the worldsheet. In this paper we would like to explore the idea of modifying the problem, by taking the level $k$ in \eqref{wzw} to be general, and studying the limit $k\to\infty$. This idea is in the spirit of general large $N$ expansions in QFT. As is well known, it often happens that large $N$ theories capture qualitatively, and sometimes even quantitatively, features of the theory one is actually interested in, which has $N$ of order one. At the same time, at large $N$ many theories become more amenable to analytic treatment.

Of course, without the factor of ${\mathbb{R}}^d$ in \eqref{wzw}, the level $k$ WZW CFT has been studied in a $1/k$ expansion already in the original paper \cite{Witten:1983ar}. In that regime, one can think about the WZW CFT as a sigma model on a large three-sphere with $k$ units of $H$-flux. It was shown in \cite{Witten:1983ar} that the resulting theory is conformal for a particular radius of the sphere, of order $\sqrt k l_s$. Moreover, it was shown that many features of the theory at general $k$ are visible at large $k$. The hope is that this is still the case in our problem, where the ${\mathbb{R}}^d$ is present, and the couplings $\phi_{a\bar b}$ depend on the radial coordinate in this space. 

In the next sections we will perform the large $k$ calculation of the effective action, but before turning to it we would like to make some preliminary remarks that will help organize the calculation. 

First, looking at \eqref{su2k}, \eqref{compsu2l}, we see that at large $k$ it is convenient to rescale the currents, such that their two point function is equal to one. Thus, we define new currents, and, by abuse of notation,  denote them by the same letter. These currents satisfy the OPE
\begin{equation}
\label{rescaledsu2k}
    J^a(z)J^b(0)\sim \frac{\delta^{ab}}{z^2}+\alpha\frac{i{\epsilon^{ab}}_c}{z}J^c(0)\ ,
\end{equation}
where 
\begin{equation}
\label{defal}
    \alpha=\sqrt{\frac{2}{k}}
\end{equation}
is a parameter that we will take to be small. We will take the currents in the perturbation \eqref{lwsgen} to be the rescaled ones \eqref{rescaledsu2k} rather than the original ones \eqref{su2k}, i.e.
\begin{equation}
\label{lwsgenres}
    {\cal L}_{\rm int}=-\phi_{a\bar b}(x)J^a\bar J^{\bar b}~,
\end{equation}
and study the resulting theory to leading order in $1/k$ but for arbitrary perturbations $\phi_{a\bar b}$.  

To see how the large $k$ limit simplifies the calculation, let's consider the different terms in the effective action on ${\mathbb{R}}^d$ in the derivative expansion. The term with no derivatives is the potential for the $\phi_{a\bar b}$'s. The first thing to note about this potential is that it vanishes at $k=\infty$. Indeed, in this limit $\alpha\to 0$ in \eqref{defal}, and the non-abelian affine Lie algebra \eqref{rescaledsu2k} becomes abelian. Thus, the non-abelian Thirring deformation \eqref{lwsgenres} becomes abelian. As is well known, the abelian Thirring model has a vanishing $\beta$-function. The $\phi_{a\bar b}$ parametrize in this case a Narain-type moduli space. In the spacetime language, this is equivalent to the potential vanishing at $\alpha=0$. It is not hard to see that the leading contribution to the potential goes like $\alpha^2\sim 1/k$. Below we will calculate it exactly as a function of $\phi_{a\bar b}$.

The next terms in the derivative expansion of the effective action are the two derivative terms. These take the form 
\begin{equation}
\label{metric}
    L_K=G^{a\bar b,c\bar d}(\phi_{i\bar j})\nabla\phi_{a\bar b}\nabla\phi_{c\bar d}\ .
\end{equation}
The metric $G^{a\bar b,c\bar d}(\phi_{i\bar j})$ is non-trivial in the limit $\alpha\to 0$. It is essentially the Zamolodchikov metric on the Narain moduli space labeled by the $\phi$'s,
\ie
\label{zammet}
G^{a\bar b,c\bar d}=(2\pi)^2|z|^4\langle J^a(z) \bar J^{\bar b}(\bar z) J^c (0)\bar J^{\bar d}(0)\rangle~.
\fe
We will compute it in the next section. 

To summarize the discussion so far, the effective Lagrangian up to two derivatives takes the form 
\ie
\label{efflag}
L_{\rm eff}=G^{a\bar b,c\bar d}(\phi_{i\bar j})\nabla\phi_{a\bar b}\nabla\phi_{c\bar d}+\alpha^2 V(\phi_{a\bar b})\ .
\fe
Both the metric $G$ and potential $V$ in \eqref{efflag} are of order $\alpha^0$. The factor of $\alpha^2$ in the potential discussed above is exhibited explicitly in \eqref{efflag}. 

We next demonstrate that to leading order in $1/k$ we can neglect all the corrections to 
\eqref{efflag}. To see that, we note that the Lagrangian \eqref{efflag} has the following property. If we rescale the coordinates on ${\mathbb{R}}^d$ by a factor of $\alpha$, i.e. define new coordinates 
\ie
\label{rescxy}
y^i=\alpha x^i,\; i=1,\cdots, d
\fe
the relative factor $\alpha^2$ in \eqref{efflag} disappears, and we find a Lagrangian without any small parameters. If the solutions of the e.o.m. of this Lagrangian vary on a scale of order one in $y$, in terms of $x$ they vary on the scale
$1/\alpha$. This scale becomes arbitrarily large in the large $k$ limit. 

To study such solutions, we can neglect all the corrections to the Lagrangian \eqref{efflag}. For example, higher derivative corrections to this Lagrangian are smaller by powers of $\alpha$, due to the fact that the $\phi_{a\bar b}$'s depend on the variable $y$ \eqref{rescxy}. Terms in the potential that scale like higher powers of $\alpha$ can also be neglected in the small $\alpha$ limit.  

The second simplification associated with the large $k$ limit has to do with the potential term in \eqref{efflag}. Earlier in this section we argued that calculating the potential to all orders in the fields $\phi_{a\bar b}$ is difficult because of the necessity to subtract massless exchanges before taking the momenta to zero in S-matrix elements. However, it is clear from that discussion that these difficulties afflict terms of higher order in $\alpha$. Indeed, \eqref{efflag} implies that all the vertices in the EFT scale like $\alpha^2$ (or higher powers of $\alpha$). Since massless exchanges involve at least two such vertices, the subtleties in taking the zero momentum limit discussed above have to do with terms in the effective action that go like $\alpha^4$ or higher. Therefore, to leading order in $1/k$ the calculation simplifies. 

In addition to the fields $\phi_{a\bar b}$ we need to add to the discussion the $d$ dimensional dilaton field $\Phi$ and metric $g$. As usual, \cite{Polchinski:1998rq}, their inclusion modifies the Lagrangian \eqref{efflag} to  
\ie
\label{efflagdil}
L_{\rm eff}=\sqrt{g}e^{-2\Phi}\left[-\mathcal{R}-4(\nabla\Phi)^2+G^{a\bar b,c\bar d}(\phi_{i\bar j})\nabla\phi_{a\bar b}\nabla\phi_{c\bar d}+\alpha^2 V(\phi_{a\bar b})\right]\ .
\fe
Here, $\mathcal{R}$ denotes the scalar curvature corresponding to the metric $g$. After rescaling $x$, as in \eqref{rescxy}, the Lagrangian becomes 
\ie
\label{resclagdil}
L_{\rm eff}=\alpha^{2-d}\sqrt{g}e^{-2\Phi}\left[-\mathcal{R}-4(\nabla\Phi)^2+G^{a\bar b,c\bar d}(\phi_{i\bar j})\nabla\phi_{a\bar b}\nabla\phi_{c\bar d}+ V(\phi_{a\bar b})\right]
\fe
where the fields depend on $y$, \eqref{rescxy}, and the derivatives are all w.r.t. this coordinate. The terms with more derivatives of $\Phi$ can be neglected as before, and the potential $V(\phi_{a\bar b})$ does not depend on $\Phi$, since we are studying the classical theory. 

Note also that we assumed that the kinetic terms for the metric and dilaton in \eqref{efflagdil} do not depend on $\phi_{a\bar b}$ (to leading order in $\alpha$). To justify this, consider for example, the dilaton kinetic term. We can multiply the $(\nabla\Phi)^2$ in this equation by an arbitrary function $F(\phi_{a\bar b})$, and ask what are the constraints on this function. Clearly, it must go to one when all $\phi_{a\bar b}\to 0$. It must also be $SU(2)_L\times SU(2)_R$ invariant. 
Another constraint on the function $F(\phi_{a\bar b})$ is due to the following observation. 

One can turn on a  dilaton that is linear in the coordinates on ${\mathbb{R}}^d$, say $\Phi=Q_1x^1$. The resulting worldsheet theory is obviously still a CFT, for any $Q_1$. It has the form \eqref{wzw}, with the ${\mathbb{R}}^d$ factor replaced by a linear dilaton CFT \cite{Polchinski:1998rq}. The central charge of this CFT depends on $Q_1$ in a simple way. From the point of view of the spacetime effective action \eqref{efflagdil}, the  linear dilaton theory is an exact solution of the e.o.m. of the Lagrangian \eqref{efflagdil}, with a constant potential that depends on $Q_1$. 

In that theory, the term $F(\phi_{a\bar b})(\nabla\Phi)^2$ gives rise to a potential proportional to $F(\phi_{a\bar b})$ for the fields $\phi_{a\bar b}$. However, we know that at $\alpha=0$ this potential must vanish. The reason is that, as before, in this limit the non-abelian algebra \eqref{rescaledsu2k} abelianizes, and the perturbations \eqref{lwsgenres} become standard Narain moduli. Therefore, we conclude that $F(\phi_{a\bar b})$ must be independent of $\phi_{a\bar b}$, i.e. $F=1$ for all $\phi_{a\bar b}$, to leading order in $\alpha$. 

To summarize, in this section we showed that the Lagrangian \eqref{efflagdil}, \eqref{resclagdil} is exact (in fields and derivatives) to leading order in the $1/k$ expansion. Furthermore, we expect to be able to compute exactly the metric $G^{a\bar b,c\bar d}(\phi_{i\bar j})$ and potential $V(\phi_{a\bar b})$, and use them to study generalizations of the HP solutions in $d>6$ dimensions. Motivated by these arguments, we next turn to calculating the metric \eqref{zammet} and potential $V$ in \eqref{efflag}. 

\section{The metric}
\label{metric}

To compute the metric $G^{a\bar b,c\bar d}(\phi_{i\bar j})$ (\ref{zammet}), we need to calculate the two point function 
\ie
\label{twopt}
\langle J^a(z) \bar J^{\bar b}(\bar z) J^c (0)\bar J^{\bar d}(0)\rangle~,
\fe
at a finite value of the couplings $\phi_{a\bar b}$ \eqref{lwsgenres}, but at $\alpha=0$. This can be done by expanding \eqref{twopt} in a power series in the interaction, and computing each term using the OPE algebra \eqref{rescaledsu2k} (with $\alpha=0$).  

Since at \(\alpha=0\) the OPE algebra \eqref{rescaledsu2k} becomes abelian, the two-point function \eqref{twopt} is equivalent to the well known Zamolodchikov metric on the Narain moduli space of \(\mathbb{T}^3\). In the present context, it was calculated in~\cite{Kutasov:1989dt} for the special case \(\phi_{a\bar b}=\phi\,\delta_{a\bar b}\), and more generally in~\cite{Sagkrioti:2018abh}.\footnote{Although the worldsheet theory studied in~\cite{Sagkrioti:2018abh} differs slightly from ours, the computation of \eqref{twopt} is essentially identical.} For completeness, we will include it here.

As is well known \cite{Kutasov:1988xb}, this calculation is sensitive to the choice of contact terms between the left and right-moving currents,
\ie
\label{contterms}
\langle J^a(z) \bar J^{\bar b}(\bar w) \rangle = C^{a\bar b}\delta^2(z-w)~.
\fe
These contact terms have a natural geometric interpretation in the space of field theories. They are related to the Christoffel symbols of the Zamolodchikov metric \eqref{zammet} at a particular point in this space. Since we will be expanding around the point $\phi_{a\bar b}=0$, we are interested in the contact terms \eqref{contterms} at that point. 

In general, contact terms are not part of the CFT data, and we can choose them to take any value. Changing the contact terms corresponds to a reparametrization of the space of theories. In our case there is a natural choice. At the point $\phi_{a\bar b}=0$, the theory has an $SU(2)_L\times SU(2)_R$ symmetry. The contact terms \eqref{contterms} break that symmetry for any non-zero $C^{a\bar b}$, since they transform as $(3,3)$ under it. Therefore, it is natural to set them to zero, to preserve the symmetry. We will make that choice below. 

We are now ready to return to the calculation of the two point function \eqref{twopt}. Expanding it in a power series in $\phi_{a\bar b}$, we have 
\begin{equation}
\label{abcd}
\begin{split}
   \frac{G^{a\bar b,c\bar d}(\phi_{i\bar j})}{(2\pi)^2|z|^4}=\sum_{n=0}^\infty\frac{1}{n!}\left(\prod_{i=1}^n\phi_{a_i\bar b_i}\int d^2z_i\right) \Big\langle J^a(z)\bar J^{\bar b}(\bar z)J^c(0)\bar J^{\bar d}( 0)\prod_{i=1}^nJ^{a_i}(z_i)\bar J^{\bar b_i}(\bar z_i)\Big\rangle&\\
   =\sum_{n=0}^\infty\frac{1}{n!}\left(\prod_{i=1}^n\phi_{a_i\bar b_i}\int d^2z_i\right) \Big\langle J^a(z)J^c(0) \prod_{i=1}^nJ^{a_i}(z_i)\Big\rangle\Big\langle \bar J^{\bar b}(\bar z)\bar J^{\bar d}(0)\prod_{i=1}^n\bar J^{\bar b_i}(\bar z_i)\Big\rangle&\ .    
\end{split}
\end{equation}
The $n=0$ term in \eqref{abcd} is given by the two-point function in the undeformed theory, 
\begin{equation}
\label{G0}
    G^{a\bar b,c\bar d}_0(\phi_{i\bar j})=(2\pi)^2|z|^4\big\langle J^a(z)J^c(0)\big\rangle\big\langle \bar J^{\bar b}(\bar z)\bar J^{\bar d}(0)\big\rangle=(2\pi)^2\delta^{ac}\delta^{\bar b\bar d}\ .
\end{equation}
With the choice of contact terms described above, all the terms with odd $n$ in \eqref{abcd} vanish, while the ones with even $n$ can be computed using Wick contractions, with the propagator \eqref{rescaledsu2k} at $\alpha=0$. 

For the terms with $n\ge2$ in (\ref{abcd}), we start by using Wick contractions to eliminate $J^a(z)$. There are two choices:
$J^a(z)$ can contract with either $J^c(0)$ or with one of the $J^{a_i}(z_i)$. As a result,  the $n$'th term in the sum  (\ref{abcd}), $G^{a\bar b,c\bar d}_{n}(\phi_{i\bar j})$, is given by
\begin{equation}
\label{Gn}
    G^{a\bar b,c\bar d}_{n}(\phi_{i\bar j})=G^{a\bar b,c\bar d}_{n,1}(\phi_{i\bar j})+G^{a\bar b,c\bar d}_{n,2}(\phi_{i\bar j})\ ,
\end{equation}
where
\begin{equation}
\label{Gn1}
    G^{a\bar b,c\bar d}_{n,1}(\phi_{i\bar j})=(2\pi)^2\frac{\bar z^2\delta^{ac}}{n!}\left(\prod_{i=1}^n\phi_{a_i\bar b_i}\int d^2z_i\right) \Big\langle \prod_{i=1}^nJ^{a_i}(z_i)\Big\rangle\Big\langle \bar J^{\bar b}(\bar z)\bar J^{\bar d}( 0)\prod_{i=1}^n\bar J^{\bar b_i}(\bar z_i)\Big\rangle\ ,
\end{equation}
and
\begin{equation}
\label{Gn2}
  G^{a\bar b,c\bar d}_{n,2}(\phi_{i\bar j})=\frac{(2\pi)^2}{(n-1)!}\left(\prod_{i=1}^n\phi_{a_i\bar b_i}\int d^2z_i\right)\frac{|z|^4\delta^{aa_n}}{(z-z_n)^2}\\
  \Big\langle J^c(0)\prod_{i=1}^{n-1}J^{a_i}(z_i)\Big\rangle\Big\langle\bar J^{\bar b}(\bar z)\bar J^{\bar d}(0)\prod_{i=1}^n\bar J^{\bar b_i}(\bar z_i)\Big\rangle\ .
\end{equation}
In (\ref{Gn1}), $\bar J^{\bar b}(\bar z)$ must contract with one of the $\bar J^{\bar b_i}(\bar z_i)$'s to get a connected diagram.\footnote{Note that \eqref{Gn1} is  the $n$'th term in the expansion of $\langle \bar J^{\bar b}(\bar z) \bar J^{\bar d}(0)\rangle$.} Performing this contraction, we find 
\begin{equation}
G^{a\bar b,c\bar d}_{n,1}(\phi_{i\bar j})=\frac{(2\pi)^2\bar z^2\delta^{ac}}{(n-1)!}\left(\prod_{i=1}^n\phi_{a_i\bar b_i}\int d^2z_i\right) \frac{\delta^{\bar b\bar b_n}}{(\bar z-\bar z_n)^2}\Big\langle \prod_{i=1}^nJ^{a_i}(z_i)\Big\rangle\Big\langle \bar J^{\bar d}( 0)\prod_{i=1}^{n-1}\bar J^{\bar b_i}(\bar z_i)\Big\rangle\ .
\end{equation}
Contracting $J^{a_n}(z_n)$ with one of the other $J^{a_i}(z_i)$'s gives
\begin{equation}
\label{Gn11}
\begin{split}
G^{a\bar b,c\bar d}_{n,1}(\phi_{i\bar j})=&(2\pi)^2\frac{\bar z^2\delta^{ac}}{(n-2)!}\left(\prod_{i=1}^n\phi_{a_i\bar b_i}\int d^2z_i\right) \frac{\delta^{a_{n-1}a_n}\delta^{\bar b\bar b_n}}{(z_n-z_{n-1})^2(\bar z-\bar z_n)^2}\\
&\Big\langle \prod_{i=1}^{n-2}J^{a_i}(z_i)\Big\rangle\Big\langle \bar J^{\bar d}( 0)\prod_{i=1}^{n-1}\bar J^{\bar b_i}(\bar z_i)\Big\rangle\ .
\end{split}
\end{equation}
Integrating (\ref{Gn11}) over $z_n$ using equation (\ref{dz22}) gives
\begin{equation}
\begin{split}
G^{a\bar b,c\bar d}_{n,1}(\phi_{i\bar j})=&(2\pi)^2\frac{\bar z^2\delta^{ac}\phi^{a_{n-1}\bar b}}{(n-2)!}\left(\prod_{i=1}^{n-1}\phi_{a_i\bar b_i}\int d^2z_i\right) (2\pi)^2\delta^{(2)}(z-z_{n-1})\\
&\Big\langle \prod_{i=1}^{n-2}J^{a_i}(z_i)\Big\rangle\Big\langle \bar J^{\bar d}( 0)\prod_{i=1}^{n-1}\bar J^{\bar b_i}(\bar z_i)\Big\rangle\ .
\end{split}
\end{equation}
Further integrating over $z_{n-1}$, we find 
\begin{equation}
\label{reGn1}
\begin{split}
G^{a\bar b,c\bar d}_{n,1}(\phi_{i\bar j})=&(2\pi)^4\frac{\bar z^2\delta^{ac}}{(n-2)!}\phi^{a_{n-1}\bar b}\phi_{a_{n-1}\bar b_{n-1}}\left(\prod_{i=1}^{n-2}\phi_{a_i\bar b_i}\int d^2z_i\right) \\
&\Big\langle\prod_{i=1}^{n-2}J^{a_i}(z_i)\Big\rangle\Big\langle \bar J^{\bar b_{n-1}}(\bar z)\bar J^{\bar d}( 0)\prod_{i=1}^{n-2}\bar J^{\bar b_i}(\bar z_i)\Big\rangle\\
=&(2\pi)^2\phi^{e\bar b}\phi_{e\bar f}G^{a\bar f,c\bar d}_{n-2,1}(\phi_{i\bar j})\ .
\end{split}
\end{equation}
Note that for $n=2$, this formula still applies, with
\begin{equation}
\label{G01}
    G^{a\bar b,c\bar d}_{0,1}(\phi_{i\bar j})=(2\pi)^2 \delta^{ac}\delta^{\bar b\bar d}\ ,
\end{equation}
which coincides with $G^{a\bar b,c\bar d}_0(\phi_{i\bar j})$ (\ref{G0}). This is compatible with \eqref{Gn}, since $G^{a\bar b,c\bar d}_{0,2}(\phi_{i\bar j})=0$.

Summing both sides of the recursion relation (\ref{reGn1}) over $n$, using (\ref{G01}), we find
\begin{equation}
    \sum_{n=2}^\infty G^{a\bar b,c\bar d}_{n,1}(\phi_{i\bar j})=(2\pi)^2\phi^{e\bar b}\phi_{e\bar f}\sum_{n=4}^\infty G^{a\bar f,c\bar d}_{n-2,1}(\phi_{i\bar j})+(2\pi)^4\delta^{ac}\phi^{e\bar b}{\phi_{e}}^{\bar d}\ .
\end{equation}
So,
\begin{equation}
\label{sGn1}
    \sum_{n=2}^\infty G^{a\bar b,c\bar d}_{n,1}(\phi_{i\bar j})=(2\pi)^4({\delta^{\bar f}}_{\bar b}-4\pi^2\phi^{e\bar f}\phi_{e\bar b})^{-1}\delta^{ac}\phi^{e\bar f}{\phi_{e}}^{\bar d}\ ,
\end{equation}
where $(\cdots)^{-1}$ denotes the $(\bar b\bar f)$ matrix element of the inverse of the $3\times 3$ matrix in brackets.

We next turn to (\ref{Gn2}). In this term, $J^{a_n}(z_n)$ has already been eliminated by Wick contraction. Thus, it is useful to contract $\bar J^{\bar b_n}(\bar z_n)$ as well. Its contraction with $\bar J^{\bar b}(\bar z)$ gives a formally divergent contribution to the one point function  $\langle J^c(0)\bar J^{\bar d}(0)\rangle$, which vanishes by conformal invariance (see also appendix \ref{secint}). The contraction with $\bar J^{\bar d}(0)$ gives a contribution to the contact term \eqref{contterms} (after the integration over $z_n$ using (\ref{dz22})), which again is not what we want in \eqref{abcd}. 

Therefore, to get a contribution to \eqref{Gn2}, we need to contract $\bar J^{\bar b_n}(\bar z_n)$ with one of the other $\bar J^{\bar b_i}(\bar z_i)$'s. There are $n-1$ equivalent choices of such a contraction. Together they lead to
\begin{equation}
\begin{split}
G^{a\bar b,c\bar d}_{n,2}(\phi_{i\bar j})=&(2\pi)^2\frac{|z|^4\delta^{aa_n}\delta^{\bar b_{n-1}\bar b_n}}{(n-2)!}\left(\prod_{i=1}^n\phi_{a_i\bar b_i}\int d^2z_i\right)\frac{1}{(z-z_n)^2(\bar z_n-\bar z_{n-1})^2}\\
&\Big\langle J^c(0)\prod_{i=1}^{n-1}J^{a_i}(z_i)\Big\rangle\Big\langle \bar J^{\bar b}(\bar z)\bar J^{\bar d}(0)\prod_{i=1}^{n-2}\bar J^{\bar b_i}(\bar z_i)\Big\rangle \ .
\end{split}
\end{equation}
Integrating over $z_n$ (using (\ref{dz22})) and $z_{n-1}$, we find
\begin{equation}
\label{Gn21}
\begin{split}
G^{a\bar b,c\bar d}_{n,2}(\phi_{i\bar j})
=&(2\pi)^4\frac{|z|^4\phi^{a\bar f}\phi_{e\bar f}}{(n-2)!}\left(\prod_{i=1}^{n-2}\phi_{a_i\bar b_i}\int d^2z_i\right)\\
&\Big\langle J^e(z)J^c(0)\prod_{i=1}^{n-2}J^{a_i}(z_i)\Big\rangle\Big\langle\bar J^{\bar b}(\bar z)\bar J^{\bar d}(0)\prod_{i=1}^{n-2}\bar J^{\bar b_i}(\bar z_i)\Big\rangle\\
=&(2\pi)^2\phi^{a\bar f}\phi_{e\bar f} G^{e\bar b,c\bar d}_{n-2}(\phi_{i\bar j})\ .
\end{split}
\end{equation}
Thus, we find that 
\begin{equation}
\label{sGn2}
   \sum_{n=2}^\infty G^{a\bar b,c\bar d}_{n,2}(\phi_{i\bar j})=(2\pi)^2\phi^{a\bar f}\phi_{e\bar f} \sum_{n=0}^\infty G^{e\bar b,c\bar d}_n(\phi_{i\bar j})\
   =(2\pi)^2\phi^{a\bar f}\phi_{e\bar f}G^{e\bar b,c\bar d}(\phi_{i\bar j})\ .
\end{equation}

\noindent
Combining the partial results (\ref{G0}), (\ref{sGn1}) and (\ref{sGn2}), we have
\begin{equation}
\begin{split}
   G^{a\bar b,c\bar d}(\phi_{i\bar j})=(2\pi)^2(\delta_{\bar b\bar d}-4\pi^2{\phi^{e}}_{\bar b}\phi_{e\bar d})^{-1}\delta^{ac}+(2\pi)^2\phi^{a\bar f}\phi_{e\bar f}G^{e\bar b,c\bar d}(\phi_{i\bar j})\ ,
\end{split}
\end{equation}
or, equivalently,
\begin{equation}
\label{Gphi}
    G^{a\bar b,c\bar d}(\phi_{i\bar j})=(2\pi)^2(\delta_{ac}-4\pi^2\phi_{a\bar f}{\phi_c}^{\bar f})^{-1}(\delta_{\bar b\bar d}-4\pi^2{\phi^{e}}_{\bar b}\phi_{e\bar d})^{-1}\ .
\end{equation}

\noindent
As discussed above, we are interested in the case where $\phi_{a\bar b}$ takes the form (\ref{lws}), where as mentioned in section \ref{seclargek} (above (\ref{rescaledsu2k})), the currents are rescaled. For $a,\bar b=1,2,3$, we have
\begin{equation}
\label{phiab}
    \phi_{a\bar b}=\begin{pmatrix}
        -\frac{1}{\sqrt{2}}{\rm Re}\;\chi&-\frac{1}{\sqrt{2}}{\rm Im}\;\chi&0\\
        \frac{1}{\sqrt{2}}{\rm Im}\;\chi&-\frac{1}{\sqrt{2}}{\rm Re}\;\chi&0\\
        0&0&\phi
    \end{pmatrix}\ .
\end{equation}
Note that for $\chi=-\sqrt{2}\phi$, the matrix \eqref{phiab} is equal to $\phi I_3$, where $I_3$ is the $3\times3$ identity matrix. In this case, a diagonal subgroup of $SU(2)_L\times SU(2)_R$ is preserved by the deformation \eqref{lwsgenres}. This is the case studied in \cite{Kutasov:1989dt} from the Thirring perspective. 

In terms of \eqref{phiab}, the matrix in (\ref{Gphi}) can be written as
\begin{equation}
    \delta_{ac}-4\pi^2\phi_{a\bar f}{\phi_c}^{\bar f}=\begin{pmatrix}
        1-2\pi^2|\chi|^2&0&0\\
        0&1-2\pi^2|\chi|^2&0\\
        0&0&1-4\pi^2\phi^2
    \end{pmatrix}\ .
\end{equation}
Inverting it and substituting into (\ref{Gphi}) gives
\begin{equation}
\label{Gchiphi}
     G^{a\bar b,c\bar d}(\phi_{i\bar j})=\frac{(2\pi)^2\delta^{ac}\delta^{\bar b\bar d}}{(1-4\pi^2\lambda^2_a)(1-4\pi^2\lambda^2_{\bar b})}\ ,
\end{equation}
where $\lambda_1=\lambda_2=|\chi|/\sqrt 2$ and $\lambda_3=\phi$.

With the metric (\ref{Gchiphi}), the kinetic term in (\ref{efflag}) takes the form
\begin{equation}
\label{kinterm}
    L_K=\frac{(2\pi)^2|\nabla \chi|^2}{(1-2\pi^2|\chi|^2)^2}+\frac{(2\pi)^2(\nabla \phi)^2}{(1-4\pi^2\phi^2)^2}\ .
\end{equation}
As a check, for $\chi=-\sqrt2\phi$, \eqref{kinterm} reduces to the expression found in~\cite{Kutasov:1989dt}, while for general $\chi$, $\phi$ it agrees with~\cite{Sagkrioti:2018abh}.

From the kinetic term \eqref{kinterm} we can read off the metric on $\phi$-space,
\ie
\label{metphi}
ds^2=\frac{(2\pi)^2d\phi^2}{(1-4\pi^2\phi^2)^2}\ .
\fe
We can change coordinates on this space, to
\begin{equation}
\label{defphitilde}
    \tilde\phi=\frac{1}{2}\ln \frac{1+2\pi\phi}{1-2\pi\phi}\ ,\quad -1<2\pi\phi<1\ ,
\end{equation}
such that the metric \eqref{metphi} becomes $ds^2=d\tilde\phi^2$, and the kinetic term for $\phi$ becomes $(\nabla\tilde \phi)^2\subset L_K$. Note that for small $\phi$, $\tilde \phi=2\pi\phi+O(\phi^2)$. 

The metric on the complex $\chi$ plane,
\begin{equation}
    ds^2=\frac{(2\pi)^2d\chi d\bar\chi}{(1-2\pi^2|\chi|^2)^2}
\end{equation}
is curved, and therefore it cannot be simplified in a similar way. Of course, particular lines in this two dimensional space, such as the line ${\rm Im}\;\chi=0$ can be treated similarly to $\phi$ above.

\section{The potential}
\label{potential}

In this section, we calculate the potential $V(\phi_{a\bar b})$ in (\ref{efflag}). Formally, this potential is given by the partition sum of the worldsheet CFT at a given value of the couplings $\phi_{a\bar b}$ in \eqref{lwsgenres},
\begin{equation}
\label{defv}
    \alpha^2V(\phi_{a\bar b})\sim \left\langle e^{-\int d^2z \mathcal{L}_{\rm int}}\right\rangle=\left\langle e^{\int d^2z \phi_{a\bar b}J^a\bar J^{\bar b}}\right\rangle\ .
\end{equation}
The fact that we are interested in the potential means that we take the 
worldsheet couplings (or, equivalently, spacetime fields) $\phi_{a\bar b}$ to be independent of position in ${\mathbb{R}}^d$. The $\sim$ in \eqref{defv} has to do with the familiar fact that to get the potential we need to divide the partition sum by the volume of the $SL(2,C)$ Conformal Killing Group (CKG) of the sphere (see e.g. \cite{Kutasov:1989dt} for a discussion in a closely related context).

Consider, for example, the leading term in the expansion of \eqref{defv} in a Taylor series in $\phi_{a\bar b}$, which is cubic in the fields,
\begin{equation}
\label{vthree}
    \alpha^2V_3\sim\frac16\prod_{j=1}^3\phi_{a_j\bar b_j}\int d^2z_j
    \left\langle \prod_{i=1}^3J^{a_i}(z_i)\bar J^{\bar b_i}(\bar z_i)
    \right\rangle\ .
\end{equation}
Using \eqref{rescaledsu2k}, \eqref{vthree} takes the form 
\begin{equation}
\label{vthreea}
V_3\sim-\frac16\phi_{a_1\bar b_1}\phi_{a_2\bar b_2}\phi_{a_3\bar b_3}
\epsilon^{a_1a_2a_3}\epsilon^{\bar b_1\bar b_2\bar b_3}    
\prod_{j=1}^3\int d^2z_j\frac{1}{|z_{12}|^2|z_{13}|^2|z_{23}|^2}\ .
\end{equation}
The last factor in \eqref{vthreea} is the volume of the CKG, $\Omega$. We have to divide by it to get the cubic term in the potential,
\begin{equation}
\label{vthreefinal}
V_3=-\frac16\phi_{a_1\bar b_1}\phi_{a_2\bar b_2}\phi_{a_3\bar b_3}
\epsilon^{a_1a_2a_3}\epsilon^{\bar b_1\bar b_2\bar b_3} =-\det\phi_{a\bar b}~.   
\end{equation}
In equation \eqref{vthreea} we omitted a dimensionful multiplicative factor, of the form $Cm_s^2$ in front of the potential $V(\phi_{a\bar b})$. We will continue omitting it for now, and will restore it later in the paper. The dimensionless constant $C$ is universal -- it does not depend on the particular worldsheet theory we are studying, and in particular it does not depend on the level $k$ of the $SU(2)$ current algebra. We will verify this fact below.  

Note that \eqref{vthreefinal} is exact (in $\alpha$) -- it does not rely on a small $\alpha$ (or large $k$, \eqref{defal}) approximation. At higher orders in $\phi_{a\bar b}$, the situation is more complicated. As discussed in section \ref{seclargek}, there are in general some additional divergences that complicate the calculation of the potential \eqref{defv}. However, as explained there, these divergences appear in subleading orders in $\alpha$, and therefore will not influence our analysis. 

The above discussion can be summarized by the formula  
\begin{equation}
\label{precv}
    \alpha^2V(\phi_{a\bar b})=\Omega^{-1}\left\langle e^{\int d^2z \phi_{a\bar b}J^a\bar J^{\bar b}}\right\rangle\ .
\end{equation}
Our goal in the rest of this section is to calculate $V(\phi_{a\bar b})$ to leading order in $\alpha$ (i.e. to order $\alpha^0$) but exactly in $\phi_{a\bar b}$. 

We will actually calculate the first derivative of the potential w.r.t. $\phi_{a\bar b}$. This quantity is useful both from the spacetime and worldsheet points of view. In spacetime, it enters the Euler-Lagrange equations of the Lagrangian \eqref{efflag}, \eqref{efflagdil}. On the worldsheet, it is directly related to the $\beta$-function of the deformed CFT \eqref{lwsgenres}, via the gradient flow relation 
\ie
\label{gradflow}
\alpha^2
\frac{\partial V}{\partial\phi_{a\bar b}}=G^{a\bar b,c\bar d}\beta_{c\bar d}~,
\fe
where $G^{a\bar b,c\bar d}$ is the metric computed in section \ref{metric}, and $\beta_{c\bar d}$ the $\beta$-function of the 
non-abelian Thirring model \eqref{lwsgenres}.  

Thus, we start with
\begin{equation}
    \alpha^2\frac{\partial V}{\partial \phi_{a\bar b}}=\Omega^{-1}\int d^2z\Big\langle J^a(z)J^{\bar b}(\bar z)e^{\phi_{c\bar d}\int d^2wJ^c(w)\bar J^{\bar d}(\bar w)}\Big\rangle\  ,   
\end{equation}
and expand it in a power series in $\phi_{a\bar b}$, as in (\ref{abcd}),
\begin{equation}
\label{Vab}
\begin{split}
    \alpha^2\frac{\partial V}{\partial \phi_{a\bar b}}
    =&\Omega^{-1}\int d^2z\sum_{n=0}^\infty\frac{1}{n!}\left(\prod_{i=1}^n\phi_{a_i\bar b_i}\int d^2z_i\right)\Big\langle J^a(z)\prod_{i=1}^nJ^{a_i}(z_i)\Big\rangle\Big\langle J^{\bar b}(\bar z) \prod_{i=1}^n\bar J^{\bar b_i}(\bar z_i)\Big\rangle\ .    
\end{split}
\end{equation}
For the special case $\phi_{a\bar b}=\lambda\delta_{a\bar b}$, (\ref{Vab}) was calculated in~\cite{Kutasov:1989dt}. Our goal is to extend this calculation to general $\phi_{a\bar b}$. 

The first step is to use the Ward identity  \eqref{rescaledsu2k}, and its right-moving analog, to eliminate $J^a(z)$ and $\bar J^{\bar b}(\bar z)$. For the left-movers, we have 
\begin{equation}
\label{wd}
\Big\langle J^a(z)\prod_{i=1}^nJ^{a_i}(z_i)\Big\rangle\\
    =\sum_{i=1}^n\left(\frac{\delta^{aa_i}}{(z-z_i)^2}\Big\langle\prod_{j\neq i}^n J^{a_j}(z_j)\Big\rangle+\alpha\frac{ i{\epsilon^{aa_i}}_b}{z-z_i}\Big\langle J^b(z_i)\prod_{j\neq i}^n J^{a_j}(z_j)\Big\rangle\right)\ .
\end{equation}
Plugging \eqref{wd} and its right-moving analog into \eqref{Vab} gives four terms, corresponding to the choice of single or double pole for the left and right-movers. We will denote the orders of the poles by a subscript $(ij)$, with $i,j=1,2$ corresponding to single and double poles, for the left and right-movers, respectively. We next compute the four contributions in turn, starting with the $(22)$ one.   

We have
\begin{equation}
\label{V22z}
\begin{split}
    \alpha^2\left(\frac{\partial V}{\partial \phi_{a\bar b}}\right)_{22}=&\Omega^{-1}\int d^2z\sum_{n=0}^\infty\frac{1}{n!}\left(\prod_{i=1}^n\phi_{a_i\bar b_i}\int d^2z_i\right)\frac{n(n-1)\delta^{aa_n}\delta^{\bar b\bar b_{n-1}}}{(z-z_n)^2(\bar z-\bar z_{n-1})^2}\\
    &\Big\langle \prod_{i=1}^{n-1}J^{a_i}(z_i)\Big\rangle\Big\langle \bar J^{\bar b_n}(\bar z_n)\prod_{i=1}^{n-2}\bar J^{\bar b_i}(\bar z_i)\Big\rangle\ .    
\end{split}
\end{equation}
Note that we contracted $J^a(z)$ and $\bar J^{\bar b}(\bar z)$ with currents located at different positions. The contribution of the term where they are contracted with currents at the same position vanishes (see appendix \ref{secint}). 

Performing the $z$ integral in (\ref{V22z}) (using equation \eqref{dz22}) gives
\begin{equation}
\begin{split}
    \alpha^2\left(\frac{\partial V}{\partial \phi_{a\bar b}}\right)_{22}=&\Omega^{-1}\sum_{n=2}^\infty\frac{\delta^{aa_n}\delta^{\bar b\bar b_{n-1}}}{(n-2)!}\left(\prod_{i=1}^n\phi_{a_i\bar b_i}\int d^2z_i\right)(2\pi)^2\delta^2(z_n-z_{n-1})\\
    &\Big\langle \prod_{i=1}^{n-1}J^{a_i}(z_i)\Big\rangle\Big\langle \bar J^{\bar b_n}(\bar z_n)\prod_{i=1}^{n-2}\bar J^{\bar b_i}(\bar z_i)\Big\rangle\ .
\end{split}
\end{equation}
Integrating over $z_n$, relabeling indices that are summed over,  and renaming $n\to n+2$, we find
\begin{equation}
\label{V22}
\begin{split}  
\left(\frac{\partial V}{\partial \phi_{a\bar b}}\right)_{22}=&\frac{(2\pi)^2}{\alpha^2}{\phi^a}_{\bar d}{\phi_c}^{\bar b}\Omega^{-1}\sum_{n=0}^\infty\frac{1}{n!}\left(\prod_{i=1}^n\phi_{a_i\bar b_i}\int d^2z_i\right)\int d^2z\\
    &\Big\langle J^c(z) \prod_{i=1}^nJ^{a_i}(z_i)\Big\rangle\Big\langle \bar J^{\bar d}(\bar z)\prod_{i=1}^n\bar J^{\bar b_i}(\bar z_i)\Big\rangle\\
=&(2\pi)^2{\phi^a}_{\bar d}{\phi_c}^{\bar b}\frac{\partial V}{\partial \phi_{c\bar d}}\ .
\end{split}
\end{equation}

\noindent
Next, we turn to the $(11)$ contribution to (\ref{Vab}), which comes from the single pole terms in the Ward identities (\ref{wd}) for both $J^a(z)$ and $\bar J^{\bar b}(\bar z)$:
\begin{equation}
\begin{split}
    \left(\frac{\partial V}{\partial \phi_{a\bar b}}\right)_{11}=&-\Omega^{-1}\int d^2z\sum_{n=0}^\infty\frac{1}{n!}\left(\prod_{i=1}^n\phi_{a_i\bar b_i}\int d^2z_i\right)\frac{n(n-1){\epsilon^{aa_n}}_c{\epsilon^{\bar b\bar b_{n-1}}}_{\bar d}}{(z-z_n)(\bar z-\bar z_{n-1})}\\
    &\Big\langle J^c(z_n)\prod_{i=1}^{n-1}J^{a_i}(z_i)\Big\rangle\Big\langle \bar J^{\bar d}(\bar z_{n-1})\prod_{i=1}^{n-2}\bar J^{\bar b_i}(\bar z_i)\bar J^{\bar b_n}(\bar z_n)\Big\rangle\ .    
\end{split}
\end{equation}
Integrating over $z$, using  (\ref{dz11}), leads to
\begin{equation}
\label{V11}
\begin{split}
    \left(\frac{\partial V}{\partial \phi_{a\bar b}}\right)_{11}=&-\Omega^{-1}\sum_{n=2}^\infty\frac{{\epsilon^{aa_n}}_c{\epsilon^{\bar b\bar b_{n-1}}}_{\bar d}}{(n-2)!}\left(\prod_{i=1}^n\phi_{a_i\bar b_i}\int d^2z_i\right)(-2\pi)\ln |z_n-z_{n-1}|^2\\
    &\Big\langle J^c(z_n)\prod_{i=1}^{n-1}J^{a_i}(z_i)\Big\rangle\Big\langle \bar J^{\bar d}(\bar z_{n-1})\prod_{i=1}^{n-2}\bar J^{\bar b_i}(\bar z_i)\bar J^{\bar b_n}(\bar z_n)\Big\rangle\\
    =&2\pi{\epsilon^{ae}}_c{\epsilon^{\bar b\bar h}}_{\bar d}\phi_{e\bar f}\phi_{g\bar h}\Omega^{-1}\int d^2z\int d^2w\ln |z-w|^2\\
    &\Big\langle J^g(z)\bar J^{\bar d}(\bar z)J^c(w)\bar J^{\bar f}(\bar w)e^{\phi_{i\bar j}\int d^2uJ^i(u)\bar J^{\bar j}(\bar u)}\Big\rangle\ .    
\end{split}
\end{equation}
As discussed in section \ref{seclargek}, we are interested in contributions to $V$ that go like $\alpha^0$. Hence, the last line in \eqref{V11} is computed at $\alpha=0$. This is precisely the quantity computed in section \ref{metric}; it is given by $G^{g\bar d,c\bar f}(\phi_{a\bar b})/(2\pi)^2|z-w|^4$. Plugging it into \eqref{V11}, we find the integral 
\begin{equation}
\label{Omega}
    4\pi\int d^2zd^2w \frac{\ln |z-w|^2}{|z-w|^4}
\end{equation}
which, as discussed in appendix \ref{secint}, is another representation of the volume of the 
$SL(2,C)$ CKG, $\Omega$. Hence, (\ref{V11}) reduces to
\begin{equation}
\label{V11eq}
    \left(\frac{\partial V}{\partial \phi_{a\bar b}}\right)_{11}=\frac{1}{2(2\pi)^2}{\epsilon^{ae}}_c{\epsilon^{\bar b\bar h}}_{\bar d}\phi_{e\bar f}\phi_{g\bar h}G^{g\bar d,c\bar f}(\phi_{i\bar j})\ ,
\end{equation}
where the metric $G$ is given by the expression we found in section \ref{metric}, equation \eqref{Gphi}. 

It remains to calculate the two terms $\left(\frac{\partial V}{\partial \phi_{a\bar b}}\right)_{21}$ and $\left(\frac{\partial V}{\partial \phi_{a\bar b}}\right)_{12}$, arising from the double pole terms in the Ward identity for $J^a(z)$, combined with the single pole terms in the Ward identity for $\bar J^{\bar b}(\bar z)$, or the other way around. The former reads
\begin{equation}
\begin{split}
    \alpha^2\left(\frac{\partial V}{\partial \phi_{a\bar b}}\right)_{21}=&\Omega^{-1}\int d^2z\sum_{n=0}^\infty\frac{1}{n!}\left(\prod_{i=1}^n\phi_{a_i\bar b_i}\int d^2z_i\right)\sum_{j\neq l}^n\frac{i\alpha \delta^{aa_l}{\epsilon^{\bar b\bar b_j}}_{\bar d}}{(z-z_l)^2(\bar z-\bar z_j)}\\
    &\Big\langle \prod_{i= 1}^{l-1}J^{a_i}(z_i)\prod_{i= l+1}^nJ^{a_i}(z_i)\Big\rangle\Big\langle \bar J^{\bar d}(\bar z_j)\bar J^{\bar b_l}(\bar z_l)\prod_{i\neq j,l}^n\bar J^{\bar b_i}(\bar z_i)\Big\rangle\ .    
\end{split}
\end{equation}
Integrating over $z$ using equation (\ref{dz21}), one finds
\begin{equation}
\label{V21}
\begin{split}
    \left(\frac{\partial V}{\partial \phi_{a\bar b}}\right)_{21}=2\pi{\phi^a}_{\bar f}\phi_{c\bar b'} {\epsilon^{\bar b\bar b'}}_{\bar d}F^{c\bar d\bar f}(\phi_{i\bar j})\ ,
\end{split}
\end{equation}
where
\begin{equation}
\label{Fcdf}
\begin{split}
F^{c\bar d\bar f}(\phi_{i\bar j})=&\frac{i}{\alpha\Omega}\sum_{n=0}^\infty\frac{1}{n!}\left(\prod_{i=1}^n\int d^2z_i\right)\sum_{j\neq l}^n\frac{1}{z_j-z_l}\prod_{i\neq j,l}^n\phi_{a_i\bar b_i}\\
&\Big\langle J^c(z_j)\prod_{i\neq j,l}^nJ^{a_i}(z_i)\Big\rangle\Big\langle \bar J^{\bar d}(\bar z_j)\bar J^{\bar f}(\bar z_l)\prod_{i\neq j,l}^n\bar J^{\bar b_i}(\bar z_i)\Big\rangle\ .
\end{split}
\end{equation}

In appendix \ref{secF}, we show that
\begin{equation}
\label{Fsol}
    F^{c\bar d\bar f}(\phi_{i\bar j})=\frac{1}{2(2\pi)^2}\left({\delta_{\bar f}}^{\bar f'}-4\pi^2\phi_{e\bar f}\phi^{e\bar f'} \right)^{-1}\left(2\pi{\phi_{e'}}^{\bar f'}\phi_{e''\bar h}{\epsilon^{e'e''}}_g-\phi_{g\bar f''}{\epsilon^{\bar f'\bar f''}}_{\bar h}\right)G^{c\bar d,g\bar h}(\phi_{i\bar j})\ .
\end{equation}
Similarly, we have
\begin{equation}
\label{V12}
\begin{split}
    \left(\frac{\partial V}{\partial \phi_{a\bar b}}\right)_{12}=2\pi{\phi_e}^{\bar b}\phi_{a'\bar d} {\epsilon^{aa'}}_c\bar F^{\bar dce}(\phi_{i\bar j})\ ,
\end{split}
\end{equation}
where
\begin{equation}
\label{Fdce}
\begin{split}
\bar F^{\bar dce}(\phi_{i \bar j})=&\frac{i}{\alpha\Omega}\sum_{n=0}^\infty\frac{1}{n!}\left(\prod_{i=1}^n\int d^2z_i\right)\sum_{j\neq l}^n\frac{1}{\bar z_j-\bar z_l}\prod_{i\neq j,l}^n\phi_{a_i\bar b_i}\\
&\Big\langle  J^c(z_j)J^e(z_l)\prod_{i\neq j,l}^nJ^{a_i}(z_i)\Big\rangle\Big\langle \bar J^{\bar d}(\bar z_j) \prod_{i\neq j,l}^n\bar J^{\bar b_i}(\bar z_i)\Big\rangle
\end{split}
\end{equation}
is given by an analog of (\ref{Fsol}).

Combining (\ref{V22}), (\ref{V11eq}), (\ref{V21}) and (\ref{V12}) gives
\begin{equation}
\label{Vabrec}
\begin{split}
    \frac{\partial V}{\partial \phi_{a\bar b}}=&(2\pi)^2{\phi^a}_{\bar d}{\phi_c}^{\bar b}\frac{\partial V}{\partial \phi_{c\bar d}}+2\pi{\phi^a}_{\bar f}\phi_{c\bar b'} {\epsilon^{\bar b\bar b'}}_{\bar d}F^{c\bar d\bar f}(\phi_{i\bar j})\\
    &+2\pi{\phi_e}^{\bar b}\phi_{a'\bar d} {\epsilon^{aa'}}_c\bar F^{\bar dce}(\phi_{i\bar j})+\frac{1}{2(2\pi)^2}{\epsilon^{ae}}_c{\epsilon^{\bar b\bar h}}_{\bar d}\phi_{e\bar f}\phi_{g\bar h}G^{c\bar f,g\bar d}(\phi_{i\bar j})\ ,
\end{split}
\end{equation}
with $F^{c\bar d\bar f}(\phi_{i\bar j})$ and $\bar F^{\bar dce}(\phi_{i\bar j})$ given by (\ref{Fsol}) and its anti-holomorphic counterpart.

As mentioned in section \ref{seclargek}, for application to the HP problem, we are interested in deformations \eqref{lwsgenres} that are invariant under the diagonal $U(1)$ $J^3+\bar J^3$. From the spacetime point of view, these are deformations that preserve translation invariance on the Euclidean time circle. In terms of the $3\times 3$ matrix $\phi_{a\bar b}$ this means that only $\phi_{33}$, $\phi_{+-}$, and $\phi_{-+}$ are non-zero, and the matrix $\phi_{a\bar b}$ takes the form (\ref{phiab}). Taking $a=\bar b=3$ in \eqref{Vabrec} and substituting the metric (\ref{Gchiphi}) into (\ref{Fsol}) and (\ref{Vabrec}), a tedious but straightforward calculation leads to
\begin{equation}
\label{Vpphi}
    \frac{\partial V}{\partial \phi}=-\frac{1}{2}\frac{|\chi|^2}{(1-2\pi\phi)^2(1-2\pi^2|\chi|^2)^2}~.
\end{equation}
Integrating both sides of \eqref{Vpphi}, we obtain
\begin{equation}
\label{Vf}
    V=-\frac{1}{4\pi}\frac{|\chi|^2}{(1-2\pi\phi)(1-2\pi^2|\chi|^2)^2}+f(|\chi|^2)\ ,
\end{equation}
with $f(|\chi|^2)$ a function to be determined. To determine this function, we can, e.g., take the sum of (\ref{Vabrec}) over $a=b=1$ and $a=b=2$. This gives
\begin{equation}
    \frac{\partial V}{\partial{\rm Re}\;\chi}=-\frac{1}{\sqrt 2}\left(\frac{\partial V}{\partial \phi_{11}}+\frac{\partial V}{\partial \phi_{22}}\right)\ .
\end{equation}
Again after a long calculation, we find
\begin{equation}
\label{VRechi}
    \frac{\partial V}{\partial {\rm Re}\;\chi}=-\frac{\phi+\pi|\chi|^2}{(1-2\pi^2|\chi|^2)^3(1-2\pi\phi)}{\rm Re}\;\chi\ .
\end{equation}
As a check,  for $\chi=-\sqrt{2}\phi$ (\ref{Vpphi}) and (\ref{VRechi}) give  
\begin{equation}
\label{checkcp}
    \frac{\partial V}{\partial \chi}=\frac{\sqrt2\phi^2(1+2\pi\phi)^2}{(1-4\pi^2\phi^2)^4}=-\sqrt{2}\frac{\partial V}{\partial\phi}\ .
\end{equation}
Combining \eqref{checkcp} with the kinetic term (\ref{kinterm}), one can check that 
the e.o.m. are consistent with setting $\chi=-\sqrt{2}\phi$, which is a consequence of the $SU(2)$ symmetry, as explained above.

To determine the function $f(|\chi|^2)$ in (\ref{Vf}), we differentiate (\ref{Vf}) with respect to ${\rm Re}\;\chi$, and subtract (\ref{VRechi}). This gives a differential equation for $f(|\chi|^2)$,
\begin{equation}
   2{\rm Re}\;\chi\ f'(|\chi|^2)=\frac{{\rm Re}\;\chi}{2\pi(1-2\pi^2|\chi|^2)^3}\ .
\end{equation}
Canceling ${\rm Re}\;\chi$ and integrating both sides, we find
\begin{equation}
\label{finalf}
    f(|\chi|^2)=\frac{1}{16\pi^3(1-2\pi^2|\chi|^2)^2}-\frac{1}{16\pi^3}\ .
\end{equation}
The additive constant is determined by the requirement that $V=0$ when $\chi=0$.

\begin{figure}
	\centering
	\subfigure[]{
	\begin{minipage}[t]{0.45\linewidth}
	\centering
	\includegraphics[width=2.5in]{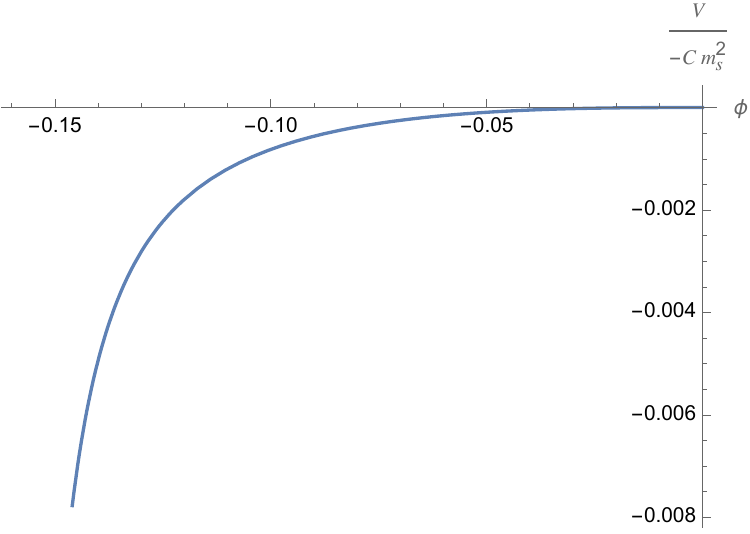}\label{Vphi}
	\end{minipage}}
	\subfigure[]{
	\begin{minipage}[t]{0.45\linewidth}
	\centering
	\includegraphics[width=2.5in]{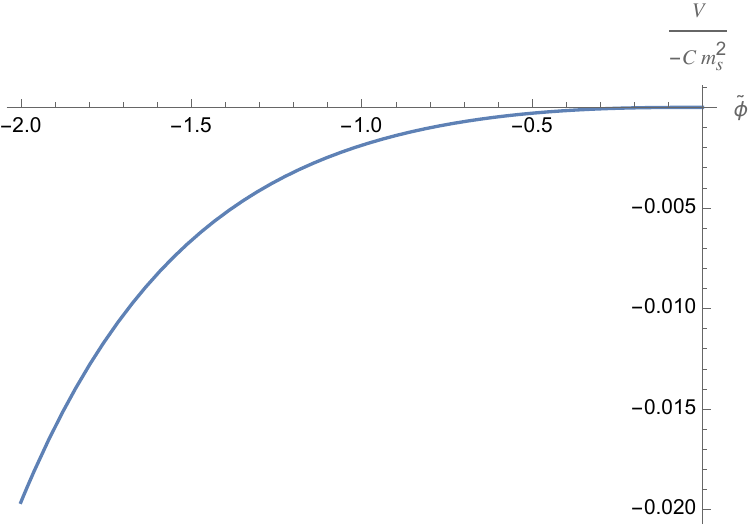}\label{Vtphi}
	\end{minipage}}
	\centering
\caption{\label{Vphitphi}The potential $V$ (\ref{Vchiphi}) for $\chi=-\sqrt{2}\phi$ as a function of (a) $\phi$, and (b) $\tilde \phi$ (\ref{defphitilde}).}
\end{figure}

Substituting \eqref{finalf} into (\ref{Vf}), we finally obtain the expression for the potential, 
\begin{equation}
\label{Vchiphi}
    V(\phi,\chi,\chi^*)=\frac{Cm_s^2}{16\pi^3}\left(-\frac{4\pi^2|\chi|^2}{(1-2\pi\phi)(1-2\pi^2|\chi|^2)^2}+\frac{1}{(1-2\pi^2|\chi|^2)^2}-1\right)~,
\end{equation}
where we have restored the overall factor $Cm_s^2$ mentioned above. This factor is computed in appendix \ref{secC}. It is given by \eqref{C27}, and in particular is negative. For $\chi=-\sqrt2\phi=\lambda/\sqrt2$, the potential \eqref{Vchiphi} agrees with the one in~\cite{Kutasov:1989dt}. In figure \ref{Vphitphi} we plot it for this case. It increases monotonically with $\phi$, or $\tilde\phi$ (\ref{defphitilde}). In terms of $\phi$ it diverges as $\phi\to-1/2\pi$, near which it behaves like $V\sim -\frac{1}{1+2\pi\phi}$. The corresponding behavior as a function of $\tilde\phi$ is $V\sim -e^{-2\tilde \phi}$ at large negative $\tilde\phi$. In addition, \eqref{Vchiphi} generalizes the calculation in~\cite{Kutasov:1989dt} to the case where $\chi$ and $-\sqrt{2}\phi$ are not equal, which in the HP context allows us to use it away from the Hagedorn temperature.

One may also verify that our results, \eqref{efflagdil}, \eqref{kinterm}, and \eqref{Vchiphi}, reproduce the Lagrangian obtained from a Kaluza--Klein reduction on \(\mathbb{S}^3\), for example the one given in eq.~(27) of~\cite{Cvetic:2000dm}\footnote{See also \cite{Sagkrioti:2018abh}.}, upon making the identification $Y=e^{\frac{8\Phi}{d-2}}$
and\footnote{We thank ChatGPT for providing this unimodular parameterization of \(\phi\) and \(\chi\).}
\begin{equation}
    \tilde T_{ij}=\begin{pmatrix}
        e^{-\tilde\phi}&0&0&0\\
        0&e^{-\tilde\phi}&0&0\\
        0&0&e^{\tilde\phi}\left(\cosh\sqrt2\,\rho+\frac{{\rm Re}\,\chi}{|\chi|}\sinh\sqrt2\,\rho\right)&-e^{\tilde\phi}\frac{{\rm Im}\,\chi}{|\chi|}\sinh\sqrt2\,\rho\\
        0&0&-e^{\tilde\phi}\frac{{\rm Im}\,\chi}{|\chi|}\sinh\sqrt2\,\rho&e^{\tilde\phi}\left(\cosh\sqrt2\,\rho-\frac{{\rm Re}\,\chi}{|\chi|}\sinh\sqrt2\,\rho\right)
    \end{pmatrix},
\end{equation}
where $\tilde\phi$ is defined in (\ref{defphitilde}) and
\begin{equation}
    \rho=\frac{1}{\sqrt2}\ln\frac{1+\sqrt2\pi|\chi|}{1-\sqrt2\pi|\chi|}\ .
\end{equation}
For small $\phi$, $\chi$, one can expand the potential \eqref{Vchiphi} in a Taylor series in the fields. The leading term in the expansion is 
\begin{equation}
\label{cubV}
    V_3=-\frac{Cm_s^2}{2}\phi|\chi|^2~,
\end{equation}
which is the analog of the cubic interaction term in \eqref{lsix}. At the next (quartic) order in the fields we have 
\begin{equation}
\label{quartV}
    V_4=-\frac{C\pi m_s^2}{4}\left(|\chi|^4+4\phi^2|\chi|^2\right)~.
\end{equation} 
Comparing it to \eqref{Leight} we see that the relative coefficient of the two terms in $V_4$ is the same here and there. As we will discuss in the next section, this is a simple consequence of the $SU(2)_L\times SU(2)_R$ symmetry of the effective Lagrangian, and must persist to all orders in the fields.

The analysis of this and the previous sections can also be used to calculate the $\beta$-function of the non-abelian Thirring model \eqref{lwsgenres} to leading order in $1/k$, using \eqref{gradflow}. It generalizes the discussion of~\cite{Kutasov:1989dt} in two ways. One is that we are not demanding that a diagonal $SU(2)$ symmetry be preserved by the deformation. The second is that we allow the couplings in the Thirring model \eqref{lwsgenres} to depend on the spatial coordinates on $\mathbb{R}^d$. In fact, the relation to the HP problem makes it clear that the interesting theories are ones where the couplings depend on position. For example, such theories allow for non-trivial fixed points of the worldsheet RG (which from the spacetime point of view are the generalized HP solutions), while for couplings that do not depend on position, there do not appear to be such fixed points (which would correspond to stationary points of the potential $V$ \eqref{Vchiphi} plotted in figure \ref{Vphitphi}).  

To summarize the discussion of sections \ref{seclargek} -- \ref{potential}, we have determined the full HP effective Lagrangian \eqref{efflagdil} to leading order in $1/k$ (or $\alpha$ \eqref{defal}). In terms of the HP fields $\phi$ and $\chi$, the kinetic term in the Lagrangian is given by \eqref{kinterm}, and the potential $V$ by \eqref{Vchiphi}. Unlike the original HP Lagrangian and its leading corrections, reviewed in section \ref{review}, our Lagrangian is valid to all orders in the fields, and can be used to study solutions in which these fields become large. We explained in section \ref{seclargek} how the large $k$ limit makes this possible. 

The discussion above makes it clear that all the results in the literature that rely on the small field approximation will be reproduced by our Lagrangian. However, our results allow one to go beyond those results (at large $k$), and construct the solutions beyond the small field approximation. The next step is to analyze these solutions. We present this analysis in the companion paper \cite{Chu:2025kzl}.

\section{Beyond the large $k$ limit}
\label{beyond}

In sections \ref{seclargek} -- \ref{potential} we studied the large $k$ limit of the HP EFT. In this section we will return to the system we are interested in, which has $k=1$ (in the bosonic string), and discuss the question how much information about the EFT we can deduce from symmetries and other considerations. We will focus mainly on the potential term in the effective Lagrangian, to demonstrate the methods, but as explained earlier, for $k\sim 1$ terms with an arbitrarily large number of derivatives are not necessarily suppressed, so the potential gives a small part of the information needed to construct generalized HP solutions for $d>6$. Nevertheless, the discussion of symmetry constraints is useful. For example, we will see that to the order studied in our previous work \cite{Balthazar:2022szl,Balthazar:2022hno}, the Lagrangian is fixed by symmetry, and that this remains the case for one more order.  

As explained in previous sections, the effective potential $V(\phi_{a\bar b})$ must be invariant under $SU(2)_L\times SU(2)_R$, even if we are interested in studying the theory away from the Hagedorn temperature. Thus, the starting point of the analysis is to classify invariants that can be constructed out of the $3\times 3$ matrix $\phi_{a\bar b}$, that transforms as $(3,3)$ under this symmetry. We next discuss the invariants that scale like $\phi_{a\bar b}^n$ for some low values of $n$. 

For $n=2$, there is a unique invariant, 
\begin{equation}
\label{twoinv}
   \phi_{a\bar b}\phi^{a\bar b}=\phi^2+|\chi|^2~,
\end{equation}
where we have used the form \eqref{phiab} for the matrix $\phi_{a\bar b}$. 
The coefficient of this invariant in the potential must vanish, since it gives rise to a mass for the radion $\phi$, which is inconsistent with the fact that this field has a flat potential. In fact, this constraint must be imposed to all orders in $\phi_{a\bar b}$. 

For $n=3$ there is again a unique $SU(2)_L\times SU(2)_R$ invariant, that we have encountered before, in equations \eqref{cubicL}, \eqref{vthreefinal}, $\det\phi_{a\bar b}$. It is clearly invariant under $SU(2)_L\times SU(2)_R$, which acts by multiplication of the matrix $\phi_{a\bar b}$ from the left and right by special unitary matrices $U_L$ and $U_R$, respectively. For $\phi_{a\bar b}$ of the form \eqref{phiab}, one has 
\begin{equation}
\label{formthree}
    \det\phi_{a\bar b}=\frac{1}{2}\phi|\chi|^2~.
\end{equation}
Thus, the $SU(2)_L\times SU(2)_R$ symmetry determines the cubic term in the potential, up to an overall coefficient (which we will discuss later). 

We next turn to the quartic terms in the potential. At this order, there are two $SU(2)_L\times SU(2)_R$ invariants one can write down. One is the square of \eqref{twoinv}, $(\phi_{a\bar b}\phi^{a\bar b})^2$. The other can be obtained by defining 
\ie
\label{defO}
O^{a\bar b}=\epsilon^{aa_1a_2}\epsilon^{\bar b\bar b_1\bar b_2}\phi_{a_1\bar b_1}\phi_{a_2\bar b_2}\ .
\fe
The $3\times 3$ matrix $O$ is quadratic in the $\phi_{a\bar b}$'s, and transforms as $(3,3)$ under $SU(2)_L\times SU(2)_R$. For $\phi_{a\bar b}$ of the form \eqref{phiab} it takes the form 
\begin{equation}
\label{matO}
    O^{a\bar b}=\begin{pmatrix}
        -\sqrt 2\phi{\rm Re}\;\chi& -\sqrt 2\phi{\rm Im}\;\chi& 0\\
        \sqrt 2\phi{\rm Im}\;\chi& -\sqrt 2\phi{\rm Re}\;\chi& 0\\
        0&0&|\chi|^2
    \end{pmatrix}\ .
\end{equation}
One can form an invariant by squaring \eqref{defO}, \eqref{matO}:
\begin{equation}
\label{fourinv}
   O_{a\bar b}O^{a\bar b}=4\phi^2|\chi|^2+|\chi|^4~.
\end{equation}

The coefficient in the effective potential of $(\phi_{a\bar b}\phi^{a\bar b})^2$ must vanish, for the same reason as before -- it gives a quartic potential to the radion $\phi$, which is inconsistent with it being a modulus. Thus, we conclude that the quartic potential for $\phi_{a\bar b}$ is completely determined, up to an overall coefficient, to be given by \eqref{fourinv}. This is consistent with the form \eqref{Leight}, that came from analyzing string scattering amplitudes, and the form \eqref{quartV} that we obtained at large $k$. Of course, this is just a consistency check on the calculations, since the $SU(2)_L\times SU(2)_R$ symmetry of the problem, that leads to the relative coefficients of the two terms in all these equations, is either implicitly or explicitly used in deriving them.  

We are now ready to discuss the constraints due to $SU(2)_L\times SU(2)_R$ invariance of the potential at higher orders in $\phi_{a\bar b}$. This matrix has $3\times 3=9$ real parameters, but $3\times 2=6$ of them are the  $SU(2)_L\times SU(2)_R$ symmetry directions, that the potential does not depend on. Thus, the potential depends on three parameters, which are invariant under the symmetries. In fact, we already constructed three such invariants, that scale as the lowest powers of the field: $\phi_{a\bar b}\phi^{a\bar b}$ \eqref{twoinv}, $\det\phi_{a\bar b}$ \eqref{formthree}, and $O_{a\bar b}O^{a\bar b}$ \eqref{fourinv}. All higher order terms are expected to be linear combinations of powers of these three invariants. 

Consider, for example the quintic terms in the potential. The only invariant we can construct out of the above building blocks is 
\begin{equation}
\label{fiveinv}
   \phi_{a\bar b}\phi^{a\bar b}\det\phi_{c\bar d}=\frac12\phi|\chi|^2(\phi^2+|\chi|^2)~.
\end{equation}
We conclude that at this order the potential is given by \eqref{fiveinv}, up to an overall coefficient, that needs to be determined using other considerations. At sixth order in the fields there are two possible structures that can appear, $(\det\phi_{a\bar b})^2$ and $\phi_{a\bar b}\phi^{a\bar b} O_{c\bar d}O^{c\bar d}$ (a third combination, $(\phi_{a\bar b}\phi^{a\bar b})^3$ is ruled out, since it gives rise to a radion potential). Therefore, at this order there are two coefficients that are not determined by the symmetries. 

Clearly, as the order in the fields increases, there are more and more coefficients that are not determined by the symmetries. The full potential $V$ can be written as
\ie
\label{Vfull}
V=\sum_{n_1, n_2, n_3=0}^\infty C_{n_1, n_2, n_3}(\phi_{a\bar b}\phi^{a\bar b})^{n_1}
(\det\phi_{c\bar d})^{n_2}(O_{e\bar f}O^{e\bar f})^{n_3}~.
\fe
Here $C_{n_1,n_2,n_3}$ are coefficients that need to be determined by other considerations. We will next discuss some constraints on these coefficients, but first we note that the large $k$ potential \eqref{Vchiphi} can indeed be written in this way, as it must, since it came from an $SU(2)_L\times SU(2)_R$ invariant expression. One can check that in terms of the invariants \eqref{twoinv}, \eqref{formthree}, \eqref{fourinv}, it takes the form 
\begin{equation}
    V(\phi,\chi,\chi^*)=-\frac{Cm_s^2}{4\pi^3}\frac{4\pi^3\det\phi_{a\bar b}-16\pi^6(\det\phi_{a\bar b})^2+\pi^4O_{a\bar b}O^{a\bar b}}{1-4\pi^2\phi_{a\bar b}\phi^{a\bar b}-64\pi^6(\det\phi_{a\bar b})^2+4\pi^4O_{a\bar b}O^{a\bar b}}\ .
\end{equation}
Coming back to the general case \eqref{Vfull}, we next discuss some constraints on the coefficients $C_{n_1,n_2,n_3}$. One such constraint is that setting the winding tachyon $\chi$ to zero, the potential for the radion $\phi$ should be flat. Looking back at the definitions \eqref{twoinv}, \eqref{formthree}, \eqref{fourinv}, we see that this implies that $C_{n_1,0,0}=0$ for all $n_1\ge0$.

We next turn to terms in \eqref{Vfull} that are quadratic in $\chi$. 
These terms are responsible for the mass of $\chi$, that depends on $\phi$. We know how that mass depends on the radius of the $\phi$ circle, \eqref{minfty}. Thus, if we know the relation between the radion $\phi$, as defined in our paper, and the radius $R$, we can obtain additional constraints on the coefficients $C_{n_1,n_2,n_3}$ in \eqref{Vfull}. We next discuss what's involved in this calculation.  

First, we note that the only terms contributing to $V$ \eqref{Vfull} that are quadratic in $\chi$,  are those with $(n_2,n_3)=(1,0)$ and $(0,1)$ (and any $n_1$). The terms with $n_2=1$ go like $\phi^{2n_1+1}$, while those with $n_3=1$ go like $\phi^{2n_1+2}$. Thus, if we know the dependence of the radius of Euclidean time $R$ on $\phi$, we can determine all the coefficients $C_{n_1,1,0}$ and $C_{n_1,0,1}$ in \eqref{Vfull}.   

To determine the relation between $R$ and $\phi$, we proceed as follows. At $R=R_H$, the Lagrangian of the field $X$ discussed in section \ref{secSU2} is 
\begin{equation}
    \mathcal{L}=\frac{1}{2\pi\alpha'}\partial X\bar \partial X\ ,\quad X\sim X+2\pi R_H\ .
\end{equation}
The field $\phi$ \eqref{lws} deforms this Lagrangian as follows:
\begin{equation}
\label{deformL}
    \mathcal{L}=\left(\frac{1}{2\pi\alpha'}+\frac{2}{\alpha'}\phi_0\right)\partial X\bar \partial X=\frac{1}{2\pi\alpha'}\partial Y\bar \partial Y\ .
\end{equation}
In \eqref{deformL}, $\phi_0$ multiplies the operator $-2 J^3\bar J^3$ (see \eqref{su2l}, \eqref{lws}). The reason for the subscript $0$ is that expressing the currents $J^3$ and $\bar J^3$ in terms of $X$ leads to a non-zero contact term \eqref{contterms}, while the way we defined $\phi$ before is with vanishing contact terms. In other words, $\phi_0$ and $\phi$ are related by a reparametrization on the space of theories \cite{Kutasov:1988xb}. 

The second equality in \eqref{deformL} helps us relate the parameter $\phi_0$ to the deformed radius $R$. For general $\phi_0$, the rescaled coordinate 
$Y=\sqrt{1+4\pi\phi_0}X$ is canonically normalized, but its identification is $Y\sim Y+2\pi R$, with 
\begin{equation}
\label{Rphi0}
    R=R_H\sqrt{1+4\pi\phi_0}\ .
\end{equation}

In order to find the relation between $R$ \eqref{Rphi0} and $\phi$, we need to find the coordinate transformation between $\phi_0$ and $\phi$. This can be done by computing the metric on the space of theories, $G=(2\pi)^2|z|^4\langle J^3(z)\bar J^3(\bar z)J^3(0)\bar J^3(0)\rangle$, in the two coordinates. 
In terms of $\phi_0$, we have 
\begin{equation}
\begin{split}
   G_0(\phi_0)=&(2\pi)^2|z-w|^4 \left\langle \frac{2}{\alpha'}\partial X\bar\partial X(z,\bar z)\ \frac{2}{\alpha'}\partial X\bar\partial X(w,\bar w)\right\rangle\\
   =&(2\pi)^2|z-w|^4\left(\frac{2}{\alpha'}\right)^2\frac{\left\langle \partial Y\bar\partial Y(z,\bar z)\partial Y\bar\partial Y(w,\bar w)\right\rangle}{(1+4\pi\phi_0)^2}\\
   =&\frac{(2\pi)^2}{(1+4\pi\phi_0)^2}~.
\end{split}
\end{equation}
In terms of $\phi$, the calculation we need to do is equivalent to one we already did in section \ref{metric}. The result is given in equation \eqref{metphi}, $G(\phi)=(2\pi)^2/(1-4\pi^2\phi^2)^2$. Equating the line elements 
$ds^2=G_0(\phi_0)d\phi_0^2=G(\phi)d\phi^2$, we find that
\begin{equation}
\label{coordtrans}
    1+4\pi\phi_0=\frac{1+2\pi\phi}{1-2\pi\phi}\ .
\end{equation}
Substituting \eqref{coordtrans} into (\ref{Rphi0}), we find
\begin{equation}
\label{Rphi}
    R=R_H\sqrt{\frac{1+2\pi\phi}{1-2\pi\phi}}=R_H e^{\tilde\phi}\ .
\end{equation}
In the second equality we used the coordinate $\tilde\phi$ introduced in \eqref{defphitilde}, in terms of which the line element (of the field space) is trivial, $ds^2=d\tilde\phi^2$. Note that this is consistent with the familiar kinetic term $(\nabla \tilde\phi)^2$ coming from the dimensional reduction of the Einstein-Hilbert action on $x_d$ with the parametrization $g_{dd}=e^{2\tilde\phi}$. The regime $R\ge R_H$ corresponds in \eqref{Rphi} to $\phi,\tilde\phi\ge 0$. When we solve the equations of motion of the EFT derived in sections \ref{seclargek} -- \ref{potential}, $\phi$ becomes a function of the radial coordinate $r$. The fact that our EFT is valid for large fields allows us to explore the (large $k$ analog of the) region where $\phi$ approaches $-1/2\pi$, where the local radius of the circle goes to zero. 

The relation between $R$ and $\phi$ \eqref{Rphi} allows us to do a number of things. First, a comparison of \eqref{Rphi} to \eqref{rx} reveals that to leading order, $\phi$ in (\ref{lsix}) is related to the one in this section by a factor of $2\pi$, namely $\phi_{\rm there}=2\pi\phi_{\rm here}+O(\phi_{\rm here}^2)$. Taking this into account, we can compare the relative coefficient between the kinetic term $|\nabla\chi|^2$ and the cubic term $\phi|\chi|^2$ in the Lagrangian \eqref{lsix} to the one in 
(\ref{efflag}), where we set $k=1$, and determine the constant $C$ in (\ref{Vchiphi}). Note that this relies on the fact that the cubic term (\ref{vthreefinal}) is independent of $k$, as mentioned before. We find
\begin{equation}
\label{C32}
    C=-\frac{16\pi^3 R_H^2}{\alpha'}\ .
\end{equation}
For the bosonic string, $R_H=2l_s$, and thus (\ref{C32}) agrees with (\ref{C27}).

The second thing we can do using \eqref{Rphi} is to calculate the mass of $\chi$, $m_\infty(\phi)$, \eqref{minfty}. From this mass, we can calculate  all the coefficients $C_{n_1,1,0}$ and $C_{n_1,0,1}$ in \eqref{Vfull}. As mentioned above, these coefficients determine the terms in the potential $V$ that go like $|\chi|^2$. Assuming that the kinetic term of $\chi$ does not depend on $\phi$, one can read off these coefficients from the Taylor expansion of $m_\infty(\phi)$. We saw in section \ref{metric} that at large $k$ this assumption about the kinetic term is correct, see equation \eqref{kinterm}. We expect this to be the case for all $k$, but have not verified this.

\section{Summary and discussion}
\label{discuss}

This paper was motivated by two questions:
\begin{itemize}
\item Can one extrapolate the $d+1$ dimensional Euclidean Schwarzschild black hole solution \eqref{bh}-\eqref{bbb} in string theory from the regime of low Hawking temperature, $\beta\gg\beta_H$, to the vicinity of the Hagedorn temperature, $\beta\sim\beta_H$?
\item What is the relation of the Horowitz-Polchinski solutions \cite{Horowitz:1997jc,Chen:2021dsw} in $d<6$ dimensions, and their generalization to $d\ge6$ \cite{Balthazar:2022hno}, to near-Hagedorn Euclidean black holes? 
\end{itemize}
In this paper we focused on the second question. As is known from previous work \cite{Horowitz:1997jc,Chen:2021dsw,Balthazar:2022hno}, which is reviewed in section \ref{review}, while for $d\le6$ the HP CFT is weakly coupled, and is thus well described by an effective field theory, the generalized HP EFT \eqref{lsix}, \eqref{Leight}, for $d>6$ it is in general strongly coupled. This can be seen by studying the theory in $d=6+\epsilon$ dimension. For small $\epsilon$, the EFT description is valid \cite{Balthazar:2022hno}, but as $\epsilon$ increases one needs to add to the effective Lagrangian terms of higher and higher order in fields and derivatives. 

The main goal of this paper was to find a weakly coupled approximation to this strongly coupled CFT. Our main idea was to utilize the underlying $SU(2)_L\times SU(2)_R$ symmetry of the HP CFT. This symmetry made an appearance in our previous work on this theory \cite{Balthazar:2022szl, Balthazar:2022hno}, where we showed that for $d>6$ the solutions at the Hagedorn temperature preserve a diagonal $SU(2)$ subgroup of $SU(2)_L\times SU(2)_R$. Below the Hagedorn temperature the symmetry is broken further, to a diagonal $U(1)$. However, the breaking is spontaneous, and the effective action should be invariant under the full $SU(2)_L\times SU(2)_R$ symmetry. 

The HP solution can be thought of as a normalizable state in the moduli space of the background \eqref{wzw}, which includes a factor of $SU(2)$ WZW at level $k=1$ (in the bosonic string). It is described by a non-abelian Thirring perturbation \eqref{lwsgenres}, with couplings $\phi_{a\bar b}$ that depend on the radial coordinate in $\mathbb{R}^d$. One can think of this dependence as a manifestation of the RG flow in the Thirring model, with the flow of the couplings as a function of the RG scale replaced by their dependence on the radial coordinate $r$. This dependence is such that the full theory is conformal. Thus, $r$ plays the role of the scale of an RG flow, reminiscent of holographic systems.  

The relation to non-abelian Thirring suggests an approach to our problem. If we change the level $k$ in \eqref{wzw} from one to a large value, the theory goes from being strongly coupled to weakly coupled. This was used in \cite{Kutasov:1989dt} to analyze the $\beta$-function of the non-abelian Thirring model to leading order in $1/k$, and in this paper we generalized that discussion to our setting, where the Thirring coupings $\phi_{a\bar b}$ depend on the radial coordinate in $\mathbb{R}^d$. Our main result is the derivation of the effective action of the model at large $k$. We showed that this action is manageable, in the sense that it only includes terms with up to two derivatives of the fields, in contrast with the situation for $k\sim 1$, where one cannot neglect terms with an arbitrary number of derivatives. Furthermore, we computed the potential and kinetic terms exactly in the fields. 

The resulting action has the property that it reproduces all the known small field solutions in the literature. At the same time, it allows one to study the large field regime in a controlled fashion, which is  currently impossible at $k\sim 1$. In particular, it allows us to study the large $k$ analogs of HP solutions for all $d$ for which such solutions exist. This involves (numerically) solving the equations of motion of the Lagrangian \eqref{efflagdil}, \eqref{kinterm}, \eqref{Vchiphi}, as a function of the dimension $d$ and temperature $T=1/\beta$, with the HP boundary conditions (that assume that the $\mathbb{S}^3$ remains regular for all $r=0$). As mentioned earlier in the paper, we describe these solutions in a companion paper \cite{Chu:2025kzl}. 

At large $k$, we expect to be able to describe our results in terms of a semiclassical picture. The background \eqref{wzw} corresponds in this picture to a sigma model on a large three-sphere, with radius $\sqrt kl_s$, supported by $B$-field flux. The non-abelian Thirring deformations \eqref{lwsgenres} correspond to deformations of the size and shape of the three-sphere, which depend on the radial coordinate  in $\mathbb{R}^d$. In particular, the winding tachyon $\chi$, which is non geometric in the $k=1$ theory, described as a CFT on $\mathbb{S}^1$, becomes a geometric mode on $\mathbb{S}^3$ at large $k$. 

In~\cite{Chu:2025kzl}, we describe the geometric deformation to first order in $\chi$ and $\phi$. As an example, we show that the deformation with $\chi=-\sqrt{2}\phi$ preserves an $SO(3)$ subgroup of the $SO(4)=SU(2)_L\times SU(2)_R$, that acts on an  $\mathbb{S}^2\subset\mathbb{S}^3$. The forms of the kinetic (\ref{kinterm}) and potential (\ref{Vchiphi}) terms in the effective Lagrangian suggest that at the critical value ($1/\sqrt2\pi$ for $\chi$ and $-1/2\pi$ for $\phi$), where the Lagrangian becomes singular, the deformed three-sphere develops a geometric singularity. We leave the analysis of this singularity to future work.

The radial size of the solutions we studied in this paper is $\sqrt k l_s$. This naturally leads to the question whether one should include in the discussion spherical harmonics on the three-sphere, which are massive, but have masses of the same order of magnitude. We believe that they are important for the dynamics, so next we comment on their role.

To describe the spherical harmonics we go back to section \ref{seclargek}. The non-abelian Thirring operators \eqref{lwsgenres} have a generalization that can be described as follows. Consider the worldsheet operators 
\begin{equation}
\label{spherical}
    \left(J\bar JV_j\right)_{j+1;m,\bar m}\ ,
\end{equation}
where the notation is as follows. $V_{j;m,\bar m}$ are primaries of the $SU(2)$ affine Lie algebra (for both the left and right-movers). Their worldsheet dimensions are 
\ie
\Delta_j=\frac{j(j+1)}{k+2};\;\;\; j=0,\frac12,1,\frac32,\cdots,\frac{k}{2}~.
\fe
The notation in \eqref{spherical} means that we couple the currents, that transform in the spin one representation of the corresponding $SU(2)$ with $V_j$, into an object that transforms in the spin $j+1$ representation of both $SU(2)$'s.

One can show that the combination \eqref{spherical} is primary under the worldsheet Virasoro. Therefore, one can associate to it a spacetime field $\phi_{j+1}$, whose mass is given by  
\begin{equation}
\label{MMjj}
    \frac{\alpha'}{4}M_j^2=\frac{j(j+1)}{k+2}~.
\end{equation}
Spacetime configurations with non-trivial profiles $\phi_{j+1}(r)$ of these fields correspond from the worldsheet point of view  to  deformations of the form  
\begin{equation}
\label{lwsjj}
    {\cal L}_{\rm int}=-\sum_j\phi_{j+1}(x)\left(J\bar JV_j\right)_{j+1}
\end{equation}
of the worldsheet Lagrangian. In \eqref{lwsjj} we suppressed the $(m,\bar m)$ indices on all the fields. Thus $\phi_{j+1}(x)$ is a $(2j+1)\times (2j+1)$ matrix of fields, which describes a normalizable perturbation of ${\mathbb{R}}^d\times \mathbb{S}^3$. Comparing \eqref{lwsjj} to \eqref{lwsgen} we see that $\phi_1(x)$ is precisely $\phi_{a\bar b}(x)$, that was the hero of our story in this paper. However, \eqref{lwsjj} includes additional fields that we have set to zero so far. Since, as discussed earlier in the paper, we are looking for solutions that preserve $J^3+\bar J^3$, only the terms in \eqref{lwsjj} that satisfy this constraint can have a non-zero expectation value.  

Naively, the fields $\phi_{j+1}$ with $j>0$ can be ignored, since they are massive, however, a closer look shows that this is not necessarily the case. Indeed, the masses \eqref{MMjj} are of order $m_s/\sqrt k$, which means that at distances much larger than $l_s\sqrt k$ the $\phi_j$ with $j>1$ decay exponentially. However, for distances of order $l_s\sqrt k$ or smaller, their contribution cannot necessarily be neglected. This distance scale is precisely the radial size of the HP solutions constructed in this paper, so we need to consider the role of the spherical harmonics more closely. 

At first sight it seems that these fields can be set to zero for the following reason. Consider a solution of the equations of motion described in this paper. Such a solution has a non-zero $\phi_1(r)$, but all the $\phi_j$ with $j>1$ are set to zero. The question is whether this is a solution of the full e.o.m. of the EFT that includes all these fields. The answer is yes, since the couplings of these fields to $\phi_1$ are at least quadratic in the fields. Consider, for example, the cubic terms in the potential. Since they involve three point functions of the vertex operators \eqref{spherical}, there are no terms in the potential of the form $\phi_1^2\phi_j$ -- the corresponding worldsheet three point function vanishes. Therefore, we conclude that the HP-type solutions studied in this paper are not modified by the addition of the massive modes \eqref{spherical} -- \eqref{lwsjj}. 

Nevertheless, we believe that the spherical harmonics play an important role in the dynamics. This belief is motivated by an analogy to fivebrane systems. We next describe this analogy, and then present a scenario for what we think happens in our case. 

As is well known, the near-horizon geometry of $k$ NS5-branes is the CHS geometry \cite{Callan:1991at},
\be
\label{NS5}
{\mathbb{R}}_\phi\times \mathbb{S}^3_k~,
\ee
where ${\mathbb{R}}_\phi$ is a linear dilaton CFT with slope $Q=\sqrt{\frac2k}$, and the three-sphere of radius $\sqrt kl_s$ is described by an $SU(2)_k$ WZW model. The background \eqref{NS5} looks similar to the one that appears in our problem, \eqref{wzw}, with the role of the radial direction in ${\mathbb{R}}^d$ played here by ${\mathbb{R}}_\phi$. An important difference between the two systems is that the worldsheet CFT \eqref{NS5} is singular, while \eqref{wzw} is regular. Indeed, in \eqref{NS5}, the string coupling goes to zero (infinity) at large (negative) positive $\phi$. As a consequence, all non-zero correlation functions in this background are singular. 

To resolve this difficulty, one can separate the fivebranes in the transverse ${\mathbb{R}}^4$. A configuration that has been extensively studied involves fivebranes equidistantly separated on a circle in the transverse space, see e.g. \cite{Kutasov:2001uf, Aharony:2004xn} for  reviews. In this background, the $SO(4)$ symmetry of rotation about the fivebranes is broken to $SO(2)\times \mathbb{Z}_k$. One can preserve an $SO(3)$ subgroup of $SO(4)$, by separating the fivebranes on a line in the transverse ${\mathbb{R}}^4$. This is analogous to the fact that in our problem, HP solutions break the $SU(2)_L\times SU(2)_R$ symmetry to $SO(2)$ below the Hagedorn temperature, and to $SO(3)$ at the Hagedorn temperature. 

Also like in our problem, the breaking is spontaneous, since it is associated with moving the fivebranes along their Coulomb branch. It is described in the bulk as a normalizable deformation of the CHS geometry \eqref{NS5}, which has a form similar to \eqref{lwsjj}, with two differences. One is that the fivebrane system is usually described in the superstring, so \eqref{lwsjj} needs to be replaced by its (worldsheet) supersymmetric analog. The other is that instead of the dependence of the coefficient functions $\phi_j(x)$ on the radial coordinate in ${\mathbb{R}}^d$, they depend on the coordinate on ${\mathbb{R}}_\phi$. We will not describe the details here (see e.g. \cite{Aharony:2004xn} for a discussion). The two features that are important to us are: (1) all the $\phi_j$ have non-zero expectation values in these solutions. These expectation values depend on the positions of the NS5-branes in the transverse ${\mathbb{R}}^4$; (2) these solutions have the property that in them the coordinate $\phi$ is bounded from below, and when it approaches the infrared bound, the three-sphere in \eqref{NS5} shrinks to zero size, creating a smooth space together with the radial direction $\phi$. More precisely, the description of this background in gravity is still singular, but the corresponding worldsheet theory is no longer singular.

Coming back to our system, $\mathbb{R}^d\times SU(2)_k$, it is natural to conjecture, by analogy, that the system we studied has two kinds of solutions. One is the solutions we discussed in this paper. In those solutions only $\phi_1(r)$ is non-zero in \eqref{lwsjj}, and its profile has the qualitative structure of HP solutions -- the three-sphere remains regular for all $r$. These solutions are described in \cite{Chu:2025kzl}. 

A second class of solutions has the property that the three-sphere in \eqref{wzw} develops a singularity at some finite value of $r$. In these solutions, all the $\phi_j$ are in general non-zero, which is necessary for imposing the boundary conditions at the tip of the geometry. It is this second class of solutions that connects smoothly to large EBH's as we vary $\beta$. We leave their study to future work.  

Another interesting direction for future work is the study of thermodynamic quantities for solutions of the equations of motion of the spacetime Lagrangian, as was done in the weak field regime in previous literature \cite{Chen:2021dsw,Balthazar:2022szl,Chu:2024ggi,Bedroya:2024igb}. To make progress, we again use (\ref{efflagdil}), (\ref{kinterm}), (\ref{Vchiphi}), as an approximation to the Lagrangian at $k=1$. As usual, 
the free energy is given by the on-shell action. The entropy can be obtained from the formula $S\sim\int d^d x(\beta\partial_\beta-1)L_{\rm eff}$, up to an overall factor $\beta/16 \pi G_N$ (where the additional $\beta$ comes from the reduction on the thermal circle).

To apply this entropy formula to our case, it is convenient to replace $\phi$ by $\tilde\phi$ using (\ref{defphitilde}), and write $\tilde\phi=\tilde\phi_{\infty}+\delta\tilde\phi$, so that the field $\delta\tilde\phi$ goes to zero at infinity. From (\ref{Rphi}), we deduce that $\beta=2\pi R_H e^{\tilde\phi_\infty}$. Expressing $\tilde\phi_\infty$ in terms of $\beta$, a simple calculation yields
\begin{equation}
    S= \frac{\beta}{16\pi G_N} \int d^dx \sqrt{g}e^{-2\Phi}\frac{\alpha^2(-C)m_s^2\beta}{8\pi\beta_H}\frac{e^{2\delta\tilde\phi}|\chi|^2}{(1-2\pi^2|\chi|^2)^2}\ .
\end{equation}
Lastly, using the free energy and entropy, one can also calculate the total mass (energy), $M=TS+F$.

Our analysis is also relevant to the study of the non-abelian Thirring model. We provided a conceptual explanation of the origin of the simplification of this model in the large $k$ limit from the spacetime perspective. Our general formula for the potential is valid beyond the original results of \cite{Kutasov:1989dt}, for general couplings $\phi_{a\bar b}$ \eqref{lwsgenres}. Indeed, plugging our results for the potential $V(\phi_{a\bar b})$ \eqref{Vabrec} and metric $G^{a\bar b,c\bar d}(\phi_{i\bar j})$ \eqref{Gphi} into (\ref{gradflow}), gives an expression for the $\beta$-functions $\beta_{a\bar b}$ for general couplings. For example, in the case where $\phi_{a\bar b}$ is a diagonal matrix, we find
\begin{equation}
\label{beta11}
    \beta_{11}=-\alpha^2Cm_s^2\frac{2\pi\phi_{11}(\phi_{22}^2+\phi_{33}^2)+\phi_{22}\phi_{33}(1+4\pi^2\phi_{11}^2)}{(2\pi)^2(1-4\pi^2\phi_{22}^2)(1-4\pi^2\phi_{33}^2)}\ ,
\end{equation}
and similar expressions for $\beta_{22}$ and $\beta_{33}$. When all the eigenvalues are equal (which is the case for $\chi=-\sqrt2\phi$ in \eqref{phiab}), \eqref{beta11} agrees\footnote{After replacing $\phi_{ii}=-\lambda/2$ (no sum over $i$), and taking into account the factor of $\pi$ in eq. (11) of \cite{Kutasov:1989dt}.} with eq. (22) in \cite{Kutasov:1989dt}. For general eigenvalues $\phi_{ii}$, \eqref{beta11} agrees with equation (6.2) in \cite{Sfetsos:2014jfa}. This agreement provides a check on both formalisms. 

More generally, the relation between the generalized Horowitz-Polchinski solutions in string theory, and the non-abelian Thirring model, is likely to have additional implications for both. We leave a further study of this relation to future work.

\section*{Acknowledgements}

We thank Ofer Aharony, Alek Bedroya, Micha Berkooz, Yiming Chen, Roberto Emparan, Sunny Itzhaki and David Wu for helpful discussions. DK thanks Tel Aviv University and the Weizmann Institute for hospitality during the conclusion of this work. This work was supported in part by DOE grant DE-SC0009924. The work of JC was further supported in part by DOE grant 5-29073.

\appendix
\section{Some useful integrals}
\label{secint}

In this appendix, we list some technical results that were used in the text. We use the conventions of \cite{Polchinski:1998rq} for the integration measure (given by eq. (2.1.7) in \cite{Polchinski:1998rq}), the definition of the delta function (2.1.8), and the form of the action for a canonically normalized scalar field, (2.1.10). 

Using these conventions, one has
\begin{equation}
\label{dz11}
    \int d^2z\frac{1}{(z-z_1)(\bar z-\bar z_2)}=-2\pi\ln|z_1-z_2|^2\ ,
\end{equation}
\begin{equation}
\label{dz21}
    \int d^2z\frac{1}{(z-z_1)^2(\bar z-\bar z_2)}=\frac{2\pi}{z_2-z_1}\ ,
\end{equation}
\begin{equation}
\label{dz22}
    \int d^2z\frac{1}{(z-z_1)^2(\bar z-\bar z_2)^2}=(2\pi)^2\delta^{(2)}(z_1-z_2)\ .
\end{equation}
Note that (\ref{dz21}) and (\ref{dz22}) can be derived from (\ref{dz11}) by taking derivatives with respect to $z_1$ and $\bar z_2$.\footnote{One may also verify (\ref{dz21}) by integrating by parts and using the identity $\partial_z\frac{1}{\bar z}=2\pi\delta^{(2)}(z)$, which follows from Stokes’ theorem.} As explained in the text, the volume of the CKG, $\Omega$, is given by the integral in (\ref{vthreea}). One can also integrate over one of the three variables and express it as: 
\begin{equation}
\label{twolog}
\Omega=\prod_{j=1}^3\int d^2z_j\frac{1}{|z_{12}|^2|z_{13}|^2|z_{23}|^2}=4\pi\int d^2z_1d^2z_2 \frac{\ln |z_{12}|^2}{|z_{12}|^4}\ ,
\end{equation}
where we used the integral~\footnote{In equations \eqref{dz11}, \eqref{twolog}, \eqref{int3pt}, we omitted for simplicity the cutoff dependence in the argument of the log.}  
\be
\label{int3pt}
\int \frac{d^2\xi} {|\xi-z|^2|\xi-w|^2}=4\pi\frac{\ln |z-w|^2}{|z-w|^2}~,
\ee
which can be derived from
\begin{equation}
\label{shapvir}
\int d^2z|z|^{2a}|w-z|^{2b}=2\pi|w|^{2(a+b+1)}
\frac{\Gamma(a+1)\Gamma(b+1)\Gamma(-a-b-1)}
{\Gamma(-a)\Gamma(-b)\Gamma(a+b+2)}\ ,
\end{equation}
by studying the limit $a,b\to -1$.\footnote{A generalization of (\ref{shapvir}), given by eq. (6.6.22) in~\cite{Polchinski:1998rq}, can also be used to check (\ref{dz11})—(\ref{dz22}).}

Another result used in the text is that the integral 
\be
\int d^2z\frac{1}{|z|^4}
\ee
can be set to zero in calculations. This is an example of the standard fact that power divergences can be regularized away in renormalizable field theory. In our context, an example of a regularization that does that is the following.

Consider the integral \eqref{shapvir} in the limit $a\to -2$, $b\to 0$. For small $b$ and fixed $a$, the r.h.s. of \eqref{shapvir} goes like 
\begin{equation}
    \frac{b}{(a+1)^2}\ .
\end{equation}
If we first send $a\to -2$, viewing $b$ as a regulator, and then send $b\to 0$, it goes to zero. 

We also encountered integrals like 
\begin{equation}
    \int d^2z\frac{1}{z^2\bar z}\ .
\end{equation}
One can treat them in a similar way, by using a generalization of \eqref{shapvir}, but in this case one can also write $z$ in polar coordinates, and observe that the integral vanishes after performing the angular integration. 

\section{Derivation of $F_{c\bar d\bar f}(\phi_{i\bar j})$}\label{secF}
Starting from the expressions of $F_{c\bar d\bar f}(\phi_{i \bar j})$ (\ref{Fcdf}) and $\bar F_{\bar dce}(\phi_{i \bar j})$ (\ref{Fdce}), we can, e.g., apply the ward identity for $\bar J^{\bar f}(\bar z_l)$ in the former as follows.
\begin{equation}
\begin{split}
     F^{c\bar d\bar f}(\phi_{i \bar j})=&\frac{i}{\alpha\Omega}\sum_{n=0}^\infty\frac{1}{n!}\left(\prod_{i=1}^n\int d^2z_i\right)\sum_{j\neq l}^n\frac{1}{z_j-z_l}\sum_{m\neq j,l}^n\phi_{a_m\bar b_m}\prod_{i\neq m,j,l}^n\phi_{a_i\bar b_i}\\
     &\Big\langle J^c(z_j)\prod_{i\neq j,l}^nJ^{a_i}(z_i)\Big\rangle\Big\langle \left(\frac{\delta^{\bar f\bar b_m}}{(\bar z_l-\bar z_m)^2}+\frac{i\alpha {\epsilon^{\bar f\bar b_m}}_{\bar h}\bar J^{\bar h}(\bar z_m)}{\bar z_l-\bar z_m}\right)\bar J^{\bar d}(\bar z_j)\prod_{i\neq j,l,m}^n\bar J^{\bar b_i}(\bar z_i)\Big\rangle\ .  
\end{split}   
\end{equation}
Integrating over $z_l$, we obtain
\begin{equation}
\label{Fcdf1}
\begin{split}
    F^{c\bar d\bar f}(\phi_{i \bar j})=&\frac{i}{\alpha\Omega}\sum_{n=0}^\infty\frac{1}{n!}\sum_{j\neq l}^n\left(\prod_{i=1}^{l-1}\int d^2z_i\right)\left(\prod_{i= l+1}^n\int d^2z_i\right)\sum_{m\neq j,l}^n\phi_{a_m\bar b_m}\prod_{i\neq m,j,l}^n\phi_{a_i\bar b_i}\\
    &\Big\langle J^c(z_j)\prod_{i\neq j,l}^nJ^{a_i}(z_i)\Big\rangle \Big\langle\Big(-2\pi\frac{\delta^{\bar f\bar b_m}}{\bar z_j-\bar z_m}+i\alpha {\epsilon^{\bar f\bar b_m}}_{\bar h}\bar J^{\bar h}(\bar z_m)2\pi\ln|z_j-z_m|^2\Big)\\
    &\bar J^{\bar d}(\bar z_j)\prod_{i\neq j,l,m}^n\bar J^{\bar b_i}(\bar z_i)\Big\rangle\ . 
\end{split}   
\end{equation}
Each term in the sum over $l$ is now explicitly independent of $l$. Therefore, we can replace this sum by an overall factor $n$. This simplifies (\ref{Fcdf1}) as
\begin{equation}
\label{Fcdf2}
\begin{split}
    F^{c\bar d\bar f}(\phi_{i \bar j})=&\frac{i}{\alpha\Omega}\sum_{n=1}^\infty\frac{1}{(n-1)!}\left(\prod_{i=1}^{n-1}\int d^2z_i\right)\sum_{j\neq m}^{n-1}\phi_{a_m\bar b_m}\prod_{i\neq j,m}^{n-1}\phi_{a_i\bar b_i}\\
    &\Big\langle J^c(z_j)J^{a_m}(z_m)\prod_{i\neq j,m}^{n-1}J^{a_i}(z_i)\Big\rangle\Big\langle \Big(-2\pi\frac{\delta^{\bar f\bar b_m}}{\bar z_j-\bar z_m}\\
    &+i\alpha {\epsilon^{\bar f\bar b_m}}_{\bar h}\bar J^{\bar h}(\bar z_m)2\pi\ln|z_j-z_m|^2\Big)\bar J^{\bar d}(\bar z_j)\prod_{i\neq j,m}^{n-1}\bar J^{\bar b_i}(\bar z_i)\Big\rangle\\
    =&\phi_{e\bar f'}\frac{i}{\alpha\Omega}\sum_{n=0}^\infty\frac{1}{n!}\left(\prod_{i=1}^n\int d^2z_i\right)\sum_{j\neq m}^n\prod_{i\neq j,m}^n\phi_{a_i\bar b_i}\Big\langle J^c(z_j) J^e(z_m)\prod_{i\neq j,m}^nJ^{a_i}(z_i)\Big\rangle\\
    &\Big\langle \bar J^{\bar d}(\bar z_j)\left(-2\pi\frac{\delta^{\bar f\bar f'}}{\bar z_j-\bar z_m}+i\alpha {\epsilon^{\bar f\bar f'}}_{\bar h}\bar J^{\bar h}(\bar z_m)2\pi\ln|z_j-z_m|^2\right)\prod_{i\neq j,m}^n\bar J^{\bar b_i}(\bar z_i)\Big\rangle\ ,
\end{split}   
\end{equation}
where in the second step we have shifted $n\to n+1$.

Comparing the two terms in (\ref{Fcdf2}) with (\ref{Fdce}) and (\ref{abcd}), we identify that
\begin{equation}
\label{FbFG}
\begin{split}
    F^{c\bar d\bar f}(\phi_{i \bar j})
    =-2\pi{\phi_e}^{\bar f}\bar F^{\bar dce}(\phi_{i \bar j})-\frac{1}{2(2\pi)^2} \phi_{e \bar f'}{\epsilon^{\bar f\bar f'}}_{\bar h}G^{c\bar d,e\bar h}(\phi_{i \bar j})\ . 
\end{split}   
\end{equation}
Similarly,
\begin{equation}
\label{bFFG}
\begin{split}
   \bar F^{\bar dce}(\phi_{i \bar j})
    =-2\pi{\phi^e}_{\bar f}F^{c\bar d\bar f}(\phi_{i \bar j})-\frac{1}{2(2\pi)^2} \phi_{e'\bar f}{\epsilon^{ee'}}_gG^{c\bar d,g\bar f}(\phi_{i \bar j})\ . 
\end{split}   
\end{equation}

With the above two equations, we can express $F(\phi_{i \bar j})$ and $\bar F(\phi_{i \bar j})$ in terms of $G(\phi_{i \bar j})$. For example, plugging (\ref{bFFG}) in (\ref{FbFG}) gives
\begin{equation}
\begin{split}
    F^{c\bar d\bar f}(\phi_{i \bar j})=&4\pi^2{\phi_e}^{\bar f}{\phi^e}_{\bar f'}F^{c\bar d\bar f'}(\phi_{i \bar j})+\frac{1}{4\pi}{\phi_e}^{\bar f}\phi_{e'\bar f'}{\epsilon^{ee'}}_gG^{c\bar d,g\bar f'}(\phi_{i \bar j})\\
    &-\frac{1}{2(2\pi)^2} \phi_{e \bar f'}{\epsilon^{\bar f\bar f'}}_{\bar h}G^{c\bar d,e\bar h}(\phi_{i \bar j})\ .
\end{split}
\end{equation}
Therefore, we solve that
\begin{equation}
    F^{c\bar d\bar f}(\phi_{i \bar j})=\frac{1}{2(2\pi)^2}\left({\delta_{\bar f}}^{\bar f'}-4\pi^2\phi_{e\bar f}\phi^{e\bar f'} \right)^{-1}\left(2\pi{\phi_{e'}}^{\bar f'}\phi_{e''\bar h}{\epsilon^{e'e''}}_g-\phi_{g\bar f''}{\epsilon^{\bar f'\bar f''}}_{\bar h}\right)G^{c\bar d,g\bar h}(\phi_{i \bar j})\ .
\end{equation}
Here $()^{-1}$ denotes the inverse of matrix.

\section{Derivation of $C$}\label{secC}
To determine the overall factor $C$ in the potential (\ref{Vchiphi}), we first note that from the coefficients of $|\nabla\chi|^2$ \eqref{kinterm} and $|\chi|^2$ \eqref{Vchiphi} in the effective action, the mass of the field $\chi$ can be written as
\be
\label{mchi1}
m_\chi^2=-\frac{C\alpha^2}{2(2\pi)^2\alpha'}\phi+O(\phi^2)\ .
\ee
To calculate $C$ in \eqref{mchi1}, we  proceed as follows. The two-point function of the vertex operators corresponding to $\chi$ and $\chi^*$, is given at a generic value of $\phi$ by 
\be
\label{2pt}
\langle J^+(z)\bar J^-(\bar z)J^-(w)\bar J^+(\bar w)\rangle_\phi=\frac{B}{|z-w|^{4h}}\ ,
\ee
where $B$ and $h$ depend on $\phi$. The former will not play a role in our discussion, but we mention in passing that the OPE (\ref{rescaledsu2k}) implies that it is equal to four at $\phi=0$. At that point the operator $J^+(z)\bar J^-(\bar z)$ has dimension $(1,1)$, i.e. $h=1$. 

Turning on $\phi$ corresponds to adding to the worldsheet Lagrangian the term $\mathcal{L}_{\rm int}=-\phi J^3\bar J^3$. This leads to a change of the dimension of the operator $J^+(z)\bar J^-(\bar z)$, which to first order in $\phi$ is 
\be
\label{hgamma}
h(\phi)=1+\gamma \phi+O(\phi^2)\ .
\ee
We can calculate $\gamma$ by evaluating \eqref{2pt} to first order in $\phi$,  
\begin{equation}
\label{bgamma}
\begin{split}
&\langle J^+(z)\bar J^-(\bar z)J^-(w)\bar J^+(\bar w)e^{\phi\int d^2\xi  J^3(\xi)\bar J^3(\bar\xi)}\rangle
=\frac{B}{|z-w|^4}\left(1-4\gamma\phi\ln|z-w|\right)~.   
\end{split}
\end{equation}
In \eqref{bgamma} we neglected the contribution of the variation of the constant $B$ with $\phi$, since it does not play a role below. 

Expanding the l.h.s. of \eqref{bgamma}, the term linear in $\phi$ is 
\begin{equation}
\label{linearphi}
\begin{split}
&\phi\int d^2\xi\langle J^3(\xi)\bar J^3(\bar \xi) J^+(z)\bar J^-(\bar z)J^-(w)\bar J^+(\bar w) \rangle\\
=&-B\alpha^2\phi\int d^2\xi\frac{1}{|\xi-z|^2|\xi-w|^2|z-w|^2}\\
=&-4\pi B\alpha^2\phi\frac{\ln|z-w|^2}{|z-w|^4}\ ,
\end{split}
\end{equation}
where in the last step we have used the integral (\ref{int3pt}). Comparing (\ref{linearphi}) with (\ref{bgamma}) gives
\begin{equation}
\label{gammapi}
    \gamma=2\pi\alpha^2\ .
\end{equation}
The variation of the dimension $h(\phi)$, \eqref{hgamma}, \eqref{gammapi}, implies a change of the mass of $\chi$, 
\be
h-\frac{\alpha'm^2_\chi}{4}=1\ .
\ee
To first order in $\phi$, we have 
\be
\label{m2chi}
m^2_\chi=\frac{4}{\alpha'}\gamma\phi=\frac{8}{\alpha'}\pi\alpha^2\phi\ .
\ee
Comparing (\ref{mchi1}) and (\ref{m2chi}) yields
\be
\label{C27}
C=-(4\pi)^3\ .
\ee
Note that it is independent of $\alpha$, \eqref{defal}, as expected \cite{Polchinski:1998rq}. 

\vskip 2cm

\bibliographystyle{JHEP}
\bibliography{HP2}

\end{document}